\documentclass[preprintnumbers,prd,floatfix,superscriptaddress,nofootinbib,notitlepage]{revtex4-1}
\pdfoutput=1
\usepackage{graphicx,epsfig}
\usepackage{amsmath}
\usepackage {amssymb}
\usepackage {longtable}
\usepackage{multirow}
\usepackage{dcolumn}
\usepackage{bm}
\usepackage{amsfonts}
\usepackage{subfigure}
\usepackage{color}
\usepackage{relsize}

\begin{document}

\title{Traversable wormholes
with double layer thin shells 
in quadratic gravity}
%\title{Wormholes with double layer thin shells in quadratic gravity}
%\title{Wormholes in quadratic gravity with double layer thin shells}

\author{Jo\~{a}o Lu\'{i}s Rosa}
\email{joaoluis92@gmail.com}
\affiliation{Institute of Physics, University of
Tartu, W. Ostwaldi 1, 50411 Tartu, Estonia}
\affiliation{University of Gda\'{n}sk, Jana Ba\.{z}y\'{n}skiego
8, 80-309 Gda\'{n}sk, Poland}

\author{Rui Andr\'{e}}
\email{rui.andre@tecnico.ulisboa.pt}
\affiliation{
Instituto de
Telecomunica\c c\~oes - IT,
Avenida Rovisco Pais 1, 1049-001 Lisboa, Portugal}

\author{Jos\'e P. S. Lemos}
\email{joselemos@ist.utl.pt}
\affiliation{Centro de Astrof\'{\i}sica e Gravita\c c\~ao  - CENTRA,
Departamento de F\'{\i}sica, Instituto Superior T\'ecnico - IST,
Universidade de Lisboa - UL, Av. Rovisco Pais 1, 1049-001
Lisboa, Portugal}

%%%%%%%%%%%%%%%%%%%%%%%%%%%%%%%%%%%%%%%%%%%%%%%%%%%%%%%%%%%%%%%%%%%%%
\begin{abstract}

In quadratic gravity, the junction conditions are six and
permit the appearance of double layer thin shells.  Double layers
arise typically in theories with dipoles, i.e., two opposite charges,
such as electromagnetic theories, and appear exceptionally in
gravitational theories, which are theories with a single charge.  We
explore this property of the existence of double layers in quadratic
gravity to find and study traversable wormholes in which the two
domains of the wormhole interior region, where the throat is located,
are matched to two vacuum domains of the exterior region via the use
of two double layer thin shells.  The quadratic gravity we use is
essentially given by a $R+\alpha R^2$ Lagrangian, where $R$ is the
Ricci scalar of the spacetime and $\alpha$ is a coupling constant,
plus a matter Lagrangian. The null energy condition, or NEC for short,
is tested for the whole wormhole spacetime. The analysis shows that
the NEC is satisfied for the stress-energy tensor of the matter in the
whole wormhole interior region, notably at the throat, and is
satisfied for some of the stress-energy tensor components of the
matter at the double layer thin shell, but is not satisfied for some
other components, namely, the double layer stress-energy distribution
component, at the thin shell.  This seems to mean that the NEC is
basically impossible, or at least very hard, to be satisfied when
double layer thin shells are present.  Single layer thin shells are
also admitted within the theory, and we present thin shell traversable
wormholes, i.e., wormholes without interior, with a
single layer thin shell at
the throat for which the corresponding stress-energy tensor
satisfies the NEC, that are asymmetric, i.e., with two different vacuum
domains of the exterior region joined at the wormhole throat.

\end{abstract}
%%%%%%%%%%%%%%%%%%%%%%%%%%%%%%%%%%%%%%%%%%%%%%%%%%%%%%%%%%%%%%%%
%\pacs{04.50.Kd,04.20.Cv,}
\maketitle

%%%%%%%%%%%%%%%%%%%%%%%%%%%%%%%%%%%%%%%%%%
\section{Introduction}\label{sec:intro}
%%%%%%%%%%%%%%%%%%%%%%%%%%%%%%%%%%%%%%%%%%%%%%%%%%%%%%%%%%%%%%%%%

Traversable wormholes are spacetime structures that connect two
otherwise distinct
domains.
In theories of gravitation, traversable wormhole solutions are ubiquitous.
In these wormhole solutions, it is imperative that within a neck that
opens up to the two antipodal mouths of each distinct domain, there is
a narrow throat without horizons. This widening of the neck at the two
sides of the throat gives rise, in geometrical terms, to the flaring out
condition,   essential to give the nontrivial topology of the
spacetime that characterizes wormholes.
In general relativity, to hold the throat and the mouths of the
wormhole against collapse, one needs matter fields that have a tension
much greater than their own energy density. Usual matter fields do not
possess these energy features, nevertheless it might happen that
exotic material made of appropriate quantum fields or some other
suitable kind of fields might provide these prerequisite elements
necessary for securing the whole wormhole structure.  To understand
the need for such exotic matter, one can resort to light spheres
moving in a wormhole background. Ingoing light spheres starting in one
of the two distinct domains enter the neck and will pass through the
throat emerging then as outgoing light spheres in the other distinct
domain of the wormhole spacetime. Thus, gravity, ordinarily an
attractive field, in the neighborhood of the throat has to be
repulsive in order to provide the necessary flaring out condition
converting converging light spheres into diverging ones.  More
specifically, in general relativity the flaring out condition entails
the violation of the null energy condition, or NEC for short, for the
matter fields.  The NEC states that the matter stress-energy tensor
$T_{ab}$ satisfies the inequality $T_{ab}k^ak^b\geq 0$, where $k^a$ is
an arbitrary null vector of the spacetime, which in turn means that,
in some specific frames of reference, the sum of the energy density
and the pressure of the matter has a nonnegative value. When the NEC
is violated, as it is the case at the wormhole throat in general
relativity, the matter is
called exotic matter.  Since all the other energy
conditions, namely, the weak, the strong, and the dominant energy
conditions, can only be satisfied if the NEC is satisfied, the NEC
operates as the basis to test the exoticity of matter. The NEC and the
other energy conditions on the matter fields are a safeguard for
classical general relativity, in that their violation raises deep
questions concerning the structure and the validity of the theory
itself. Be as it may, in general relativity
the throats of wormhole solutions are
allowed to exist only in neighborhoods where the NEC is violated.
Theories of gravity beyond general relativity, commonly called
modified theories of gravity, are generically motivated from quantum
gravity low energy phenomena or from corrections that can show up at a
cosmological scale. When applied to wormholes, these theories have
also provided mechanisms to keep the wormhole throat open. In
general, modified theories have, in their gravitational equation, the
usual curvature term of general relativity epitomized in the Einstein
tensor which depends on the metric and its first two derivatives, plus
terms that contain higher order metric curvature tensors, together
with terms for the other plausible fundamental fields, as well as a
term for the stress-energy tensor of the matter fields, in addition to
comprising all the other equations related to the new fields.  The new
terms in the modified theories of gravity can work for or against the
NEC and the other energy conditions, i.e., they can provide smaller
energy condition violations or higher ones.  Since exotic matter is
hard to find, one way out is to try to construct traversable wormhole
solutions with new fundamental terms, gravitational or otherwise, that
alleviate the NEC violation of the matter fields at the throat.
Normally, although the condition violations can be reduced, they
remain at some degree or another in the solution. Moreover, even if
one manages to have the NEC satisfied at the throat neighborhood this
does not imply that the condition cannot be violated elsewhere, and
thus other places of the wormhole spacetime may contain some form of
exotic matter.  Rarely the NEC is satisfied for the entire traversable
wormhole spacetime, from the throat to the the infinity of the two
domains.

Let us mention some traversable wormhole solutions that underline the
subtleties of their construction and in addition put some emphasis on
the energy condition violation features.  Traversable wormholes
appeared in general relativity first as solutions connecting two
different asymptotic flat spacetime domains
\cite{ellis,bronnikovi,clement}.  Their possible use for interstellar
travel was put forward in \cite{morris1}, and a firm theoretical basis
for their construction was given in
\cite{visser1,hochbergvisser}. Traversable wormholes with interesting
features connecting two distinct asymptotically de Sitter or anti-de
Sitter regions were explored in \cite{lemos1}. Wormholes supported by
phantom matter were found in \cite{sushkov2005prd}, in a cosmological
context were discussed in \cite{faraoniisrael}, in braneworld
scenarios together with their ringing properties were analyzed in
\cite{konoplyamolina}, supported by Casimir type effects were
introduced in \cite{sushkov2005cqg}, made of electrically charged
shells in a Gauss-Bonnet theory appeared in \cite{thibeault}, and built
out of shells and cosmic strings were proposed in \cite{simeone}.  In
nonminimal theories of gravity, a Wu\hskip0.06cm{}-Yang wormhole was
constructed in \cite{balakinsz}, state-of-the-art cylindrical wormhole
solutions were searched for in \cite{lemosbronnkrechet1},
a class of traversable wormholes in $f(R)$
theories of gravity, where $R$ is the Ricci scalar of the spacetime
and $f$ some function of it, was displayed in
\cite{lobooliveira}, traversable wormholes as trapped ghosts were
exhibit in \cite{bronnsush}, wormholes in
nonminimally coupled gravitational and electromagnetic fields were
carefully built in \cite{blz}, thin shell traversable wormhole
solutions in $d$-dimensional general relativity were disclosed and
classified in \cite{lemoswormhole1}, wormholes in the Brans-Dicke
theory were established in \cite{sushkov}, the structure of wormhole
throats in $f(R)$ theories was
systematized in \cite{benedictis}, wormholes
in a hybrid metric-Palatini gravity were produced in
\cite{capozziello1}, rotating wormholes with cylindrical symmetry were
developed in \cite{lemosbronnkrechet2}, empty wormhole solutions in
$f(R)$ gravity theories were attempted in \cite{criscienzo}, traversable
wormholes that minimally violate the NEC were proposed in
\cite{lopezlobomoruni}, wormhole collisions with thin shells were
featured in \cite{gao}, thin shell wormholes with double layers in a
quadratic $f(R)$ gravity were considered in \cite{eiroa1}, wormholes
with anisotropic fluid matter were built in
\cite{menchonolmorubieragarcia}, relativistic Bose-Einstein
condensates thin shell wormholes were examined in \cite{richarte},
wormholes in generalized hybrid metric-Palatini gravity, characterized
by the function $f\left(R,{\cal R}\right)$, where ${\cal R}$ is the
Ricci scalar of an additional Palatini connection field, satisfying the
matter NEC at the throat, were designed in \cite{rosa1}, a
traversable wormhole and its relation to a black bounce
was treated in
\cite{simpsonvisser}, traversable wormholes in five-dimensional
Lovelock theory were advanced in \cite{girbetcelissimeone},
spherically symmetric wormholes in a specific hybrid metric-Palatini
gravity were realized in \cite{bronnikovhybrid}, a phase space
analysis for static wormholes sustained by an exotic isotropic perfect
fluid was performed in \cite{fay}, quantum features in wormholes were
investigated in \cite{matyjasek}, the transition from a black hole to
a wormhole in braneworld scenarios was inspected in
\cite{bronnikovkonoplya}, magnetic wormholes resembling Melvin's
universe were reported in \cite{bronnikovkrechetoshurko},
Morris-Thorne wormholes in a modified $f(R,T)$ gravity, where $T$ is
the trace of the matter stress-energy tensor, were found in
\cite{chandadeypaul}, the extension of Ellis wormholes into an anti-de
Sitter background was described in \cite{salcedo}, the existence of a
wormhole in Rastall and k-essence theories was advocated in
\cite{bronnikov2021}, self colliding wormholes were obtained in
\cite{fenglemosmatznerwormholes2}, a class of wormholes with double
layers in hybrid metric-Palatini gravity was uncovered in
\cite{Rosa:2021yym}, wormholes in $f(R)$ gravity with a phantom scalar
field were identified in \cite{karakasis}, 
traversable wormholes in general relativity were contrasted
to bubble universes in \cite{lemosluz}, a generic
analysis of traversable wormholes was accomplished in \cite{kz},
traversable wormholes violating energy conditions only near the Planck
scale were revealed in \cite{maeda}, wormhole solutions in an $f(R)$
gravity were further analyzed in \cite{ali}, traversable wormholes in
higher dimensions with warps were envisaged in \cite{kar}, wormholes
with partly phantom scalar fields were implemented in
\cite{bronnikov2022}, and traversable wormhole solutions in $f(R,T)$
gravity were further examined in \cite{rosakull}.

In the investigation of traversable wormhole solutions of a given
chosen theory, a tool that is often employed is the set of junction
conditions that matches correctly one spacetime region, solution of
the theory, into another region, also solution of the theory.  Each
theory has its own set of junction conditions.  In general relativity
the full set of junction conditions was given in \cite{israel}, where
smooth matching through a boundary surface or a nonsmooth matching
through a boundary layer, i.e., a single layer thin shell, were
accomplished.  In $f(R)$ theories of gravity the set of junction
conditions were given in \cite{Deruelle:2007pt,senovilla1}.  In
quadratic theories of gravity, where $f(R)$ contains the usual Ricci
scalar term $R$ plus further terms quadratic in the curvature
invariants, the set of junction conditions was found in
\cite{senovilla2}, with the appearance, besides the smooth matching
and the single layer thin shell, of a double layer thin shell.
Moreover, junction conditions in $f\left(R,T\right)$ theories have
been worked out in \cite{Rosafrt}, junction conditions in the
generalized hybrid metric-Palatini gravity of the form $f\left(R,{\cal
R}\right)$ were inspected and displayed with applications in
\cite{rosalemos2}, and the junction conditions of Palatini
$f\left(R,T\right)$ gravity were scrutinized in
\cite{rosarubieragarcia}.

Our aim here is to find traversable wormhole solutions in a quadratic
theory of gravity making use of the double layer thin shells that
consistently occur in the theory and analyze the behavior of the NEC
throughout the whole spacetime. Double layers appear naturally from
dipole layers in theories with two opposite charges, such as
electromagnetic theories, but are rare in theories of gravitation that
have only a single charge. The quadratic theory of gravity we use has
as Lagrangian the function $f\left(R\right)=R+\alpha R^2$, with $R$
being the Ricci scalar of a spacetime endowed with a metric and
$\alpha$ being a free constant of the theory. Then, from the set of
junction conditions of generic quadratic gravity theories, we pick up
the set that fits our quadratic theory, and establish new solutions
for traversable wormholes with double layer thin shells. The NEC is
tested and it is found that it is satisfied at the throat and for the
whole wormhole interior, and at the thin shell some stress-energy
tensor components satisfy it, while others, for instance, the
double layer distribution component, do not. We also study
a class of traversable thin shell wormholes with a single layer
that satisfies the NEC.

This paper is organized as follows.  In Sec.~\ref{sec:equations}, we
introduce the quadratic theory of gravity we are going to study.  In
Sec.~\ref{sec:jcs}, we display the full set of junction conditions of
the theory, show the particular condition that yields a double layer
thin shell, indicate how this full set simplifies to the set of single
layer thin shell junction conditions and to the set of smooth boundary
surface junction conditions, and we give the full form of the
stress-energy tensor for the whole spacetime.  In
Sec.~\ref{sec:worms}, we derive a solution for a traversable wormhole
with matter in the two domains, the upper and the lower, of the
interior region that includes the throat, two double layer thin shells
at the two junction mouths, the upper and lower, and two
Schwarzschild asymptotically flat domains, the upper and lower, of
the exterior region, and test the NEC for the whole solution.  In
Sec.~\ref{sec:wormssinglelayer}, we derive a solution for an
asymmetric single layer thin shell traversable wormhole for which the
NEC is satisfied.  In Sec.~\ref{sec:concl}, we draw some conclusions.  In
Appendix~\ref{formulasanddefinitions}, we give some generalized
formulas used in the text, and in Appendix~\ref{necgeneralrelativity}
we make a plot to complement the main text.

%%%%%%%%%%%%%%%%%%%%%%%%%%%%%%%%%%%%%%%%%%%%%%%%%%%%%%%%%%%
\section{Quadratic $R+\alpha R^2$ gravity: Action and field equation}
\label{sec:equations}
%%%%%%%%%%%%%%%%%%%%%%%%%%%%%%%%%%%%%%%%%%%%%%%%%%%%%%%%%%%

Gravitational theories with the Hilbert-Einstein term $R$ in the
action, where $R$ is the Ricci scalar, plus a quadratic term $\alpha
R^2$, with $\alpha$ a coupling constant parameter, are an instance of
quadratic theories of gravity.  The general action $S$ for such
quadratic theories is
\begin{equation}\label{action}
S=\frac{1}{16\pi}\int_{\mathcal V}\,
\Big(
f\left(R\right)+{\cal L}_{\rm m}\Big) \, d{\mathcal V},
\end{equation}
in units where the constant of gravitation $G$
and the speed of light $c$ are set to one,
$\mathcal V$ is the spacetime manifold region
on which a set
of coordinates $x^a$
is defined, $d{\mathcal V}$ is the volume element defined by 
$d{\mathcal V}=\sqrt{-g}\, d^4x$,
$g$ represents
the determinant of the metric $g_{ab}$, 
$f(R)$ is given by
\begin{equation}\label{functionf}
f\left(R\right)=R+\alpha R^2,
\end{equation}
and $\mathcal L_{\rm m}$ is the matter Lagrangian which is assumed to
be minimally coupled to the metric $g_{ab}$.  The sign of the
quadratic term can be positive, zero, or negative, an interesting
situation being the cosmological inflationary Starobinski model for
which $\alpha$ is positive, proportional to the square of the inverse
of the inflaton mass.  We stick to $\alpha>0$, and put $\alpha$
proportional the square of the inverse of the Planck mass, $M_{\rm
pl}$, where here we set Planck's constant to one $\hbar=1$, and so
$M_{\rm pl}=\sqrt{\frac{\hbar c}{G}}=1$.  Equations~(\ref{action})
and~(\ref{functionf}) provide a particular case of $f(R)$ theories,
namely quadratic theories of gravity in the Ricci scalar.  One can, if
wished, include a cosmological constant $\Lambda$ term to the function
$f(R)$ in Eq.~(\ref{functionf}).  This is of greater interest in a
cosmological context, where $f\left(R\right)$ theories have given a
number of insights but, in treating wormholes generated by matter
confined to a compact region, discarding this term is a well justified
choice.

The field equations for the $f\left(R\right)$ gravity given by the
generic form of Eq.~(\ref{action}) can be obtained by taking its
variation with respect to the metric $g_{ab}$.  Denoting the
stress-energy tensor by $T_{ab}$ defined through $
T_{ab}=-\frac{2}{\sqrt{-g}}\frac{\delta(\sqrt{-g} \,{\cal L}_{\rm
m})}{\delta(g^{ab})} $, and defining $f_R\equiv \frac{d f}{d R}$, the
variation yields $ f_R R_{ab}-\frac{1}{2}g_{ab}f \left(R\right)-
\left(\nabla_a\nabla_b-g_{ab} \Box\right)f_R=8\pi T_{ab} $, where
$\nabla_a$ is the covariant derivative and $\Box=\nabla^a\nabla_a$ is
the d'Alembert operator, both written in terms of the Levi-Civita
connection $\Gamma^c_{ab}$ of the metric $g_{ab}$.  For the specific
quadratic $f\left(R\right)$ gravity theory given in
Eq.~(\ref{functionf}) one has $f_R=1+2\alpha R$, and so one obtains
the following equation of motion
\begin{eqnarray}\label{field}
\left(1+2\alpha R\right) R_{ab}-\frac{1}{2} g_{ab}\left( R+\alpha
R^2\right)- 2\alpha\left(\nabla_a\nabla_b-g_{ab} \Box\right) R =8\pi
T_{ab},
\end{eqnarray}
which is the equation that interests us here.

We could look for other actions more generic than the ones with a
quadratic $f(R)$ Lagrangian, but the appearance of a double layer thin
shell requires that $f_{RRR}=0$.  Moreover, for our purposes of
finding interesting wormhole solutions with double layer thin shells
it is appropriate to work with the specific $f(R)$ given in
Eq.~\eqref{functionf}.  Of course, single layer thin shells and smooth
boundaries traversable wormholes are also well defined in this
quadratic theory.

\section{Junction conditions for double layers, single layers,
and smooth boundary surfaces in quadratic $R+\alpha R^2$ gravity
and the stress-energy tensor $T_{ab}$}
\label{sec:jcs}

\subsection{Generic matching conditions for double layer thin shells
and the particular cases of single layer thin shells and
smooth boundary surfaces}

The junction conditions are a set of conditions that two spacetime
regions must satisfy in order to guarantee that they can be matched at
a given separation hypersurface, and thus altogether
form a whole spacetime,
solution of the field equations.

A four-dimensional spacetime region is denoted by $\mathcal V$, which
in turn is assumed to be composed of two regions, $\mathcal V^-$ and
$\mathcal V^+$ that match at a hypersurface $\Sigma$.  The metric in
the region $\mathcal V^-$ with coordinates $x^a_-$ is denoted by
$g_{ab}^-$, and the metric in the region $\mathcal V^+$ with
coordinates $x^a_+$ is denoted by $g_{ab}^+$, with latin indices
running from $0$ to $3$, $0$ being in general a time index.  In each
side of $\Sigma$, it is defined a set of coordinates $y^\alpha$, with
Greek indices running from some combination of three of the four latin
indices, where the index excluded corresponds to the direction
perpendicular to the hypersurface.  Projection vectors from the
four-dimensional regions $\mathcal V^-$ and $\mathcal V^+$ into the
three-dimensional hypersurface $\Sigma$ are defined by
$e^a_\alpha=\frac{\partial x^a}{\partial y^\alpha}$,
where $x^a$ can be either $x^a_-$ or $x^a_+$ depending which
region one is considering.  At $\Sigma$
there is a unit normal vector $n^a$ which is defined to point from
$\mathcal V^-$ to $\mathcal V^+$.  The affine parameter along the
geodesics perpendicular to $\Sigma$ is denoted by $l$, it can be a
proper time if $\Sigma$ is spacelike or a proper length if $\Sigma$ is
timelike.  The parameter $l$ is chosen to be negative in the region
$\mathcal V^-$, zero at $\Sigma$, and positive in the region $\mathcal
V^+$.  Along these geodesics perpendicular to $\Sigma$,
the infinitesimal coordinate displacement is
$dx^a=n^adl$, and the normal $n_a$ can be written as
$n_a=\epsilon \partial_a l$,
with $\epsilon=\mp1$, $-1$ for $n^a$ a timelike vector and $+1$ for
$n^a$ a spacelike vector, so that $n_a$ satisfies the normalization
condition $n^an_a=\epsilon$.  In the matching process one uses the
distribution function formalism.  For that one defines
two distribution functions, namely, the Heaviside
step function $\Theta\left(l\right)$ and the Dirac delta
function $\delta\left(l\right)$, with
$\delta\left(l\right)=\partial_l\Theta\left(l\right)$.  For
a given quantity $A$, it is possible to write
$A=A^-\Theta\left(-l\right)+A^+\Theta\left(l\right)$, where the index
$-$ indicates that the quantity $A^-$ is the value of the quantity $A$
in the region $\mathcal V^-$, and the index $+$ indicates that the
quantity $A^+$ is the value of the quantity $A$ in the region
$\mathcal V^+$.  The jump of the quantity $A$ across $\Sigma$ is
denoted by $\left[A\right]=A^+|_\Sigma-A^-|_\Sigma$.  The tangent
vector $e^a_\alpha$ and the normal $n^a$ to $\Sigma$ have zero jump by
definition, so that $\left[n^a\right]=0$ and
$\left[e^a_\alpha\right]=0$.  The junction conditions for quadratic
$f\left(R\right)$ gravity have been derived in \cite{senovilla1}.
Here, we adapt these for the quadratic gravity defined in
Eq.~\eqref{functionf}, i.e., $f\left(R\right)=R+\alpha R^2$.  There is
a total of six junction conditions for this theory.

The first junction condition of the theory
is related to the induced metric at the
matching hypersurface. The induced metric $h_{\alpha\beta}$ at
$\Sigma$ is defined
generically by $h_{\alpha\beta}=g_{ab}e^a_\alpha e^b_\beta$,
with $h_{\alpha\beta}^-$ corresponding to the induced metric from
$g_{ab}^-$, and $h_{\alpha\beta}^+$ corresponding to the induced
metric from $g_{ab}^+$.  The first junction condition is
$\left[h_{\alpha\beta}\right]=0$, i.e., the induced metric at $\Sigma$
is continuous.  Defining the induced metric with full spacetime
indices as $h_{ab}= e_a^\alpha e_b^\beta h_{\alpha\beta}$,
the first junction condition of the theory can be written as 
\begin{equation}\label{junction1}
\left[h_{ab}\right]=0\,.
\end{equation}

The second junction condition of the theory is related to the
trace of the extrinsic curvature at the
matching hypersurface from the ambient spacetime regions. 
The extrinsic curvature $K_{\alpha\beta}$
at $\Sigma$ is defined by 
$K_{\alpha\beta}=e^a_\alpha e^b_\beta\nabla_a n_b$, with
$K_{\alpha\beta}^-$ corresponding to the extrinsic curvature from the
$\mathcal V^-$ region, and $K_{\alpha\beta}^+$ corresponding to the
extrinsic curvature from the $\mathcal V^+$ region.  The trace $K$ of
the extrinsic curvature is $K^-={K^{-\alpha}}_{\alpha}$ for $\mathcal
V^-$ and $K^+={K^{+\alpha}}_{\alpha}$ for $\mathcal V^+$.  Defining
the extrinsic curvature with full spacetime indices as $K_{ab}=
e_a^\alpha e_b^\beta K_{\alpha\beta}$ and $K={K^a}_a$, the second
junction condition of the theory can be writen as
\begin{equation}\label{junction2}
\left[K\right]=0\,,
\end{equation} 
i.e., the jump in the trace of the extrinsic curvature
vanishes at the thin shell.

The third junction condition of the theory is
\begin{eqnarray}\label{stressts}
-\epsilon\left[K_{ab}\right]+
2\epsilon\alpha\left(h_{ab}n^c\left[\nabla_cR\right]-
R^\Sigma\left[K_{ab}\right]-K_{ab}^\Sigma\left[R\right]\right)
=8\pi S_{ab},
\end{eqnarray}
where $R^\Sigma$ is an average Ricci scalar at the hypersurface
$\Sigma$, defined as $R^\Sigma=\frac12\left(R^++R^-\right)$,
$K_{ab}^\Sigma$ is an average extrinsic curvature at the hypersurface
$\Sigma$, defined as
$K^\Sigma_{ab}=\frac12\left(K^+_{ab}+K^-_{ab}\right)$, and $S_{ab}$ is
the four-dimensional
surface stress-energy tensor of the matter thin shell, defined
from the surface stress-energy tensor $S_{\alpha\beta}$ of the matter
thin shell 
that lives in
$\Sigma$ as
$S_{ab}=
e_a^\alpha e_b^\beta S_{\alpha\beta}$.

The fourth junction condition of the theory is
\begin{equation}\label{s}
2\epsilon\alpha K^\Sigma\left[R\right]=8\pi P\,,
\end{equation}
where $K^\Sigma={K^{\Sigma\,a}}_a$, and 
$P$ is a quantity
that measures the external normal pressure
or tension
supported on $\Sigma$.
In general
relativity, thin shell spacetimes with nonzero
$[R]$ exist,
but
Eq.~(\ref{s}) does not manifest itself, since
$\alpha = 0$ for general relativity.

The fifth junction condition of the theory is 
\begin{equation}\label{sa}
-2\epsilon \alpha h^c_a\nabla_c\left[R\right]=8\pi F_a\,,
\end{equation}
where $F_a$ is defined as the external flux momentum.  The normal
component of it measures the normal energy flux across $\Sigma$ and
its tangential spatial components measure the
corresponding tangential stresses.  In
general relativity, thin shell spacetimes with nonzero
$\nabla_c\left[R\right]$ exist, but Eq.~(\ref{sa}) does not manifest
itself, since $\alpha = 0$ for general relativity.

The sixth junction condition of the theory
can be written quite generically as
$2\epsilon\alpha \, \Omega_{ab}\left(l\right)= 8\pi
s_{ab}\left(l\right)$, where $\Omega_{ab}(l)$ is a distributional
geometric quantity that is defined implicitly in terms of
geometrical spacetime
quantities and quantities on the layer, namely, $ \int_{\mathcal V}
\Omega_{ab}(l) Y^{ab} d{\mathcal V}=- \int_\Sigma h_{ab}\left[R\right]
n^c\nabla_cY^{ab}d\Sigma $, with $Y^{ab}$ being some arbitrary
spacetime test tensor function, $l$ being the proper distance
perpendicular to the shell defined above, and $s_{ab}(l)$ being the
stress-energy tensor distribution of the double layer that naturally
arises in this junction condition.  Note that the geometric quantity
distribution $\Omega_{ab}(l)$ and, by inheritance, the stress-energy
tensor distribution $s_{ab}(l)$, are defined through the spacetime test
tensorial function $Y^{ab}$ and, although this function is quite
arbitrary, having the sole constraint of being defined in a region of
compact support, the distributional double layer quantity
$\Omega_{ab}(l)$, and so $s_{ab}(l)$, is unique.
One can show that the
the double layer stress-energy tensor
distribution $s_{ab}\left(l\right)$
given by the sixth junction condition of the theory 
can also be written in the form, see Appendix \ref{app0},
\begin{equation}
\label{sab}
\epsilon\alpha
h_{ab}\left[R\right]n^c\nabla_c\delta\left(l\right)
=4\pi s_{ab}\left(l\right).
\end{equation}
The double layer stress-energy tensor distribution $s_{ab}(l)$ in
Eq.~(\ref{sab}) has resemblances to classical electric dipole
distributions.  Here, this dipole term arises due to the existence of
the coupling $\alpha$ and a nonzero jump in the curvature
$R$. The dipole
distribution term in Eq.~(\ref{sab}) is originated uniquely from the
quadratic term of $f(R)$ theories,
here from the term $\alpha R^2$.  Since gravitation in general does
not have different types of charge, apparently there is still no
intuitive physical interpretation for why dipole double layer
distributions should arise in such gravity theories.  In general
relativity, thin shell spacetimes with nonzero $[R]$ exist, but
Eq.~(\ref{sab}) does not manifest itself, since $\alpha = 0$ for
general relativity.

The system of equations given in Eqs.~(\ref{junction1})-(\ref{sab})
reduce to a single layer thin shell matching when the contribution of
the double layer for the thin shell vanishes.  In such a case,
$S_{ab}$ remains nonzero, while the quantities $P$, $F_a$ and
$s_{ab}\left(l\right)$ vanish.  So, the first junction condition,
Eq.~\eqref{junction1}, continues to hold, $\left[h_{ab}\right]=0$.
The second junction condition, Eq.~\eqref{junction2}, continues to
hold, $\left[ K\right]=0$.  Eq.~(\ref{s}) gives $[R]=0$, which implies
that the third junction condition, Eq.~(\ref{stressts}), is
$-\epsilon\left[K_{ab}\right]+
2\epsilon\alpha\left(h_{ab}n^c\left[\nabla_c R\right]-
R^\Sigma\left[K_{ab}\right]\right)=8\pi S_{ab}$, and the fourth
junction condition, Eq.~(\ref{s}), is $[R]=0$ itself.  The fifth
junction condition, Eq.~\eqref{sa} is trivial, since $[R]=0$ implies
$h^c_a\nabla_c\left[R\right]=0$.  The sixth junction condition,
Eq.~\eqref{sab}, is also trivial.  Then the junction conditions of the
quadratic theory of gravity given in Eq.~\eqref{field} for a matching
with a single layer thin shell are four, namely,
$\left[h_{ab}\right]=0$, $\left[K\right]=0$,
$-\epsilon\left[K_{ab}\right]+
2\epsilon\alpha\left(h_{ab}n^c\left[\nabla_c R\right]-
R^\Sigma\left[K_{ab}\right]\right)=8\pi S_{ab}$, and $[R]=0$.

The system of equations given in Eqs.~(\ref{junction1})-(\ref{sab})
reduce to a smooth matching, i.e., to a boundary surface, when both
the contributions from the double layer and the single layer thin
shell vanish.  In such a case, all the quantities $S_{ab}$, $P$,
$F_a$, and $s_{ab}\left(l\right)$ vanish.  The first junction
condition, Eq.~\eqref{junction1}, continues to hold,
$\left[h_{ab}\right]=0$.  The second junction condition,
Eq.~\eqref{junction2}, still holds, but it is not an independent
condition anymore since it is implied by the third junction condition.
This can be seen by noting that Eq.~(\ref{s}) yields $[R]=0$, which
then implies $h^c_a\nabla_c\left[R\right]=0$, and then taking the
trace of the third junction condition, Eq.~(\ref{stressts}), and using
Eq.~(\ref{junction2}), one gets $n^c\left[\nabla_cR\right]=0$, which
replacing back into Eq.~(\ref{stressts}) yields $[K_{ab}]=0$.
Finally, tracing this result one obtains Eq.~\eqref{junction2}.  So,
now instead of Eqs.~\eqref{junction2} and~\eqref{stressts}, one has
simply $[K_{ab}]=0$.  The fourth junction condition, Eq.~(\ref{s}), is
now $[R]=0$, as mentioned above.  The fifth and sixth junction
conditions, Eqs.~\eqref{sa} and \eqref{sab}, are again trivial.  Then,
the junction conditions of the quadratic theory of gravity given in
Eq.~\eqref{field} for a smooth matching, i.e., for a boundary
surface, are three, namely, $[h_{ab}]=0$, $[K_{ab}]=0$, and $[R]=0$.

\subsection{The stress-energy tensor $T_{ab}$}

The stress-energy tensor $T_{ab}$, that appears in the field equation,
Eq.~\eqref{field}, can be now given explicitly.  The quantities that
appear in it are $T_{ab}^-$ which is the stress-energy tensor in the
region $\mathcal V^-$, $T_{ab}^+$ which is the stress-energy tensor in
the region $\mathcal V^+$, and the quantities at the thin shell,
specifically, $S_{ab}$ which is the stress-energy tensor of the matter
at the thin shell, $P$ which measures the external normal pressure or
tension supported at the thin shell, $F_a$ which is the external flux
momentum at the thin shell, and $s_{ab}(l)$ which is the stress-energy
tensor distribution of the double layer  thin shell.  Then, $T_{ab}$
assumes the form
\begin{eqnarray}\label{tabquad}
T_{ab}=T_{ab}^- \Theta\left(-l\right) + T_{ab}^+\Theta\left(l\right)
+\delta\left(l\right)\left(S_{ab}+ Pn_an_b+ F_{(a}n_{b)}\right)+
s_{ab}\left(l\right)\,.
\end{eqnarray}
Note that the dipole tensor distribution $s_{ab}(l)$ of
Eq.~(\ref{sab}) is an essential piece in making sure that the
stress-energy tensor distribution is divergence free, i.e., $\nabla^a
T_{ab}=0$.

%%%%%%%%%%%%%%%%%%%%%%%%%%%%%%%%%%%%%%%%%%%%%%%%%%%%%%%%%%%%%%
\section{Traversable wormholes with a double layer thin shell
in quadratic $R+\alpha R^2$ gravity}
\label{sec:worms}
%%%%%%%%%%%%%%%%%%%%%%%%%%%%%%%%%%%%%%%%%%%%%%%%%%%%%%%%%%%%%%%%%

\subsection{General considerations}

Let us start by describing
the geometric sector of a class of
traversable wormholes with a double layer thin shell
in quadratic $R+\alpha R^2$ gravity.  The generic line element for
the four-dimensional spacetime is $ds^2=g_{ab}dx^adx^b$, where
$g_{ab}$ is the metric in the coordinate system $x^a$.  Assuming a
static spacetime and spherical coordinates
$\left(t,r,\theta,\phi\right)$, the line element can be written as
$ds^2=g_{tt}dt^2+g_{rr} dr^2+g_{\theta\theta}d\theta^2
+g_{\phi\phi}d\phi^2$,
where $g_{tt}$, 
$g_{rr}$,  $g_{\theta\theta}$,
and $g_{\phi\phi}$
are the time, radial, 
polar, and  azimuthal
metric potentials, respectively.
In spherical coordinates
$g_{\theta\theta}=r^2$ and $g_{\phi\phi}=r^2\sin^2\theta$,
and so one has 
$ds^2=g_{tt}dt^2+g_{rr} dr^2+r^2d\Omega^2$, where
$d\theta^2+\sin^2\theta d\phi^2$ has been abbreviated to
$d\Omega^2$, i.e.,
$d\Omega^2=d\theta^2+\sin^2\theta d\phi^2$ 
is the line element on the unit sphere.  For traversable wormhole
spacetimes it is convenient to write $g_{tt}=-e^{\zeta\left(r\right)}$
and $g_{rr}= \frac{1}{1- \frac{b\left(r\right)}{r}} $, where
$\zeta\left(r\right)$ is defined as the redshift function and
$b\left(r\right)$ is defined as the shape function, so that the line
element is
\begin{equation}\label{metric}
ds^2=-e^{\zeta\left(r\right)}dt^2+
\frac{dr^2}{1-
\frac{b\left(r\right)}{r}}+r^2d\Omega^2.
\end{equation}
A traversable wormhole spacetime connects two distinct spacetime
domains through a throat situated at some radius $r_0$. This implies
that for traversable wormhole spacetimes the two metric functions of
Eq.~\eqref{metric}, $\zeta\left(r\right)$ and $b\left(r\right)$,
near the throat are not arbitrary as two conditions
must be fulfilled.  First, the wormhole spacetime must have no horizons
or singularities, such that motion in the neighborhood of the wormhole
throat at $r=r_0$ is allowed, and escape to the distinct spacetime
domains is possible.  To accomplish these conditions, it is sufficient
in static spacetimes to impose that the redshift function must be
finite everywhere, or at least in some sufficiently large region
containing the throat, i.e., $|\zeta\left(r\right)|<\infty$ within
some sufficiently large $r$.  Second, the wormhole spacetime has to
satisfy the flaring out condition.  This is a geometric condition that
must be imposed at the wormhole throat guaranteeing that it is
traversable.  This condition implies that the shape function $b(r)$
must satisfy $b\left(r_0\right)=r_0$ and
$b'\left(r_0\right)<1\label{flareout}$, where here a prime denotes
derivatives with respect to $r$.
The two metric functions of Eq.~\eqref{metric},
$\zeta\left(r\right)$ and $b\left(r\right)$, also assume special forms
compatible with the asympotics of the wormhole spacetime, for instance
they can have forms that asymptotically approach
the vacuum Schwarzchild line element. 

Let us now describe the matter sector generically.
The stress-energy tensor can be written as
\begin{equation}\label{tabanigeneric}
T_{ab}=T_{ab}(r)\,,
\end{equation}
where the dependence in the radius $r$ is made explicit
so that Eqs.~\eqref{metric} and \eqref{tabanigeneric} are compatible
when one makes use of the field equation, i.e.,
Eq.~\eqref{field}.  We can
further specify the
stress-energy tensor by establishing
which type of fluid inhabits
certain given regions of the spacetime.

In searching for traversable wormhole solutions, one is also
interested in verifying whether and where the matter satisfies the NEC,
specifically, in which regions the NEC is satisfied and which pieces
of the stress-energy tensor $T_{ab}$ satisfy the NEC.
For instance, it can be shown
that for a fluid with a stress-energy tensor of the form
of an anisotropic perfect fluid,
${T_a}^b(r)=\text{diag}\left(-\rho(r),p_r(r),p_t(r),p_t(r)\right)$,
with $\rho\left(r\right)$ being the energy density,
$p_r\left(r\right)$ being the radial pressure, and
$p_t\left(r\right)$ being the transverse pressure,
the NEC, i.e., $T_{ab}k^ak^b\geq0$, where $k^a$ is
a null vector, implies that the energy density $\rho$, the radial
pressure $p_r$, and the tangential pressure $p_t$, must satisfy the
inequalities, $ \rho+p_r\geq0$ and $\rho+p_t\geq0$.
Within general relativity, the flaring out condition, i.e.,
$b\left(r_0\right)=r_0$ and $b'\left(r_0\right)<1$, and the NEC for
anisotropic perfect fluids, i.e., $\rho+p_r\geq0$ and $\rho+p_t\geq0$,
are incompatible.  This comes from the fact that the flaring out
condition, is a requirement on the geometry of the spacetime and thus
on the the Einstein tensor $G_{ab}$, which is in turn connected to the
stress-energy tensor $T_{ab}$ through the Einstein field equation
$G_{ab}= 8\pi T_{ab}$. For static and spherically symmetric spacetimes
one can show that the flaring out condition implies $\rho+p_r<0$, and
since the NEC obliges that $\rho+p_r\geq0$, we see that the two
conditions are irreconcilable, and thus traversable wormhole solutions
violate the NEC in general relativity.
However, in modified theories of gravity, the field equations can be
envisaged in the form $G_{ab}=8\pi T_{ab}^{\text{eff}}$, where
$T_{ab}^{\text{eff}}$ is an effective stress-energy tensor that
contains not only the matter stress-energy tensor $T_{ab}$, but also
the higher order curvature terms that come from the field equations,
in particular this holds for the quadratic gravity we are considering
given in Eq.~\eqref{field}. In this way of setting the problem, it is
the effective stress-energy tensor that must follow the flaring out
condition $\rho^{\text{eff}}+p_r^{\text{eff}}<0$, and thus it is
possible, in principle, that the matter stress-energy tensor $T_{ab}$
satisfies $T_{ab}k^ak^b\geq0$, i.e., $\rho+p_r\geq0$ and
$\rho+p_t\geq0$, and so satisfies the NEC, with the flaring out condition
being simultaneously satisfied, as long as the extra higher order
curvature terms provide a sufficient negative contribution to compensate
the positive matter contributions. At the very least 
modified theories, if not having traversable wormhole solutions
that satisfy the NEC, should mitigate the violations of the NEC
for the matter.

Our wormhole spacetime $\mathcal V$ is composed of three regions.
These three regions are the wormhole interior $\mathcal V^-$ which
contains two domains at each side of the throat, two boundaries
$\Sigma$, one at each mouth, and the wormhole exterior $\mathcal V^+$
which contains two domains each one starting at each mouth $\Sigma$.
The two domains belonging to the exterior region are thus connected to
each other through the mouths, the two domains of the interior region,
and the throat.  The interior region is described by a line element of
the form given in Eq.~\eqref{metric} and some specific stress-energy
tensor for Eq.~\eqref{tabanigeneric} which are valid for the two
domains that come out of each side of the throat.  The two domains of
the exterior region, which in general can be different, can also be
described by a specific form of the line element of
Eq.~\eqref{metric}.  If one considers that the domains of the exterior
region are vacuum regions, the stress-energy tensor of
Eq.~\eqref{tabanigeneric} is zero.  The two boundaries from the
interior to the exterior are dictated by the junction conditions
previously found.  At these boundaries one may have double layer thin
shells, single layer thin shells, or boundary surfaces, i.e., a smooth
matching without any thin shells.  Here, we find a traversable
wormhole with a double layer thin shell.  To study the situation of
the NEC for the matter, one must analyze it in the three regions, in
the interior region, at the shell, and in the exterior region.

\subsection{Traversable wormhole solutions with a double layer
thin shell satisfying the NEC at the throat}
\label{sec:wormsol}

\subsubsection{The interior made of two symmetric domains}
\label{sec:necthroat}

To find an interior solution with a throat we have to specify the
redshift and shape functions,
$\zeta\left(r\right)$ and $b(r)$, respectively,
that appear in Eq.~\eqref{metric}.  A
specific form for these functions is 
$\zeta\left(r\right)=\zeta_0\left(\frac{r_0}{r}\right)$ and
$b\left(r\right)=r_0\left(\frac{r_0}{r}\right)$, so that the model has
redshift and shape functions proportional to the inverse of the radius
$r$.
With this choice, the line element 
of Eq.~\eqref{metric} 
suitable for the interior
region is 
\begin{align}\label{metricconcrete}
ds^2=-{\rm e}^{\zeta_0\left(\frac{r_0}{r}\right)} dt^2+
\frac{dr^2}
{1- {\left(\frac{r_0}{r}\right)^{2}}}
+r^2d\Omega^2\,,\quad\quad\quad\quad r_0\leq r<r_\Sigma\,,
\end{align}
where we avoided to put a $-$ in all quantities, as it should be for
quantities in the $\mathcal V^-$ region, in order to not overload the
formulas with symbols.  The radial coordinate in
Eq.~\eqref{metricconcrete} ranges from the throat radius $r_0$ to
$r_\Sigma$, $r_0\leq r< r_\Sigma$, the upper domain of the
interior region
starting at $r_0$ and extending to the upper separation boundary
at $r_\Sigma$, the upper mouth, and the lower domain of the
interior region starting
at $r_0$ and extending to the lower separation boundary at
$r_\Sigma$, the lower mouth.  
One could have defined, instead, two broad families for the
redshift and shape functions that are in line with the requirements
for a traversable wormhole, as
$\zeta\left(r\right)=\zeta_0\left(\frac{r_0}{r}\right)^\gamma $ and
$b\left(r\right)=r_0\left(\frac{r_0}{r}\right)^\beta$ with $\gamma$
and $\beta$ being constant exponents. We choose to specify $\gamma=1$
and $\beta=1$, since it is sufficient to find an interesting double
layer thin shell wormhole solution.  In the
Appendix~\ref{formulasanddefinitions} we deal with
arbitrary $\gamma$ and $\beta$.

The stress-energy tensor
considered for the interior $\mathcal V^-$ is that 
of an anisotropic
perfect fluid, which 
can be written as
\begin{equation}\label{tabani}
{T_a}^b(r)=\text{diag}\left(-\rho(r),p_r(r),p_t(r),p_t(r)\right)
\,,\quad\quad\quad\quad r_0\leq r<r_\Sigma\,,
\end{equation}
where again we avoided to put a $-$ in all quantities, as it should be for
quantities in the $\mathcal V^-$ region, in order to not overload the
formulas with symbols,
$\rho=\rho\left(r\right)$ is defined as the energy density,
$p_r=p_r\left(r\right)$ is defined as the radial pressure, and
$p_t=p_t\left(r\right)$ is defined as the transverse pressure
of the matter. The
three quantities $\rho(r)$, $p_r(r)$, and $p_t(r)$, depend on the radial
coordinate alone, so that the matter distribution is static and
spherically symmetric.
The radial coordinate in
Eq.~\eqref{tabani} ranges from the throat radius $r_0$ to
$r_\Sigma$, $r_0\leq r< r_\Sigma$, the upper domain of the
interior region
starting at $r_0$ and extending to the upper separation boundary
at $r_\Sigma$, the upper mouth, and the lower domain of the
interior region starting
at $r_0$ and extending to the lower separation boundary at
$r_\Sigma$, the lower mouth.

\begin{figure}[h!]
\includegraphics[scale=0.7]{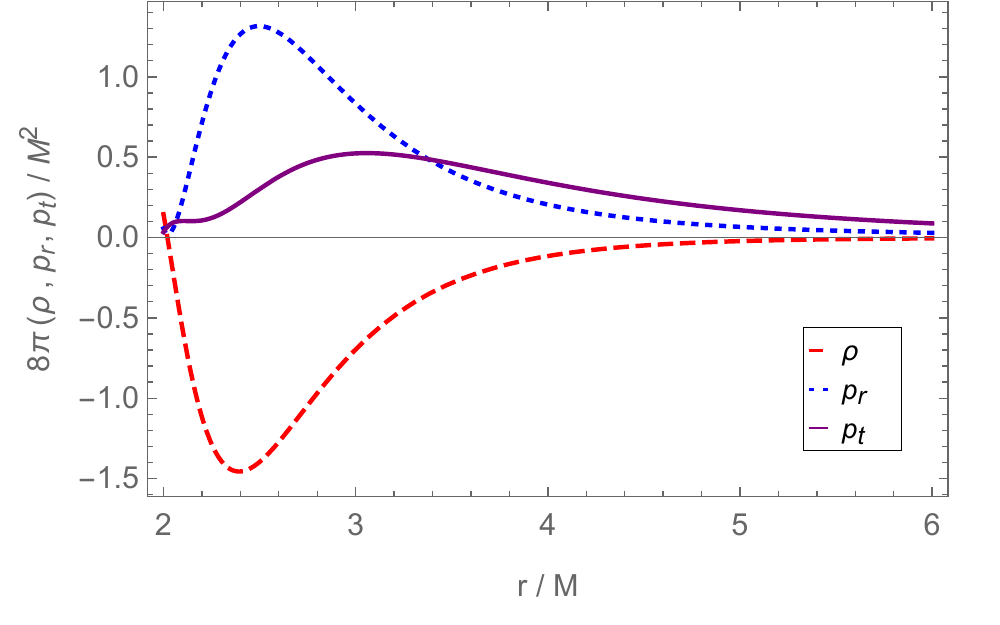}
\caption{
Plots of the
matter fields $\rho$, $p_r$, and $p_t$ as functions of
the radius $r$ for $\alpha=0.013$ in Planck units, $\zeta_0=-60$, and
$r_0=2M$, where here $M$ is some mass constant.  The
quantities plotted are normalized to $M$.}
\label{fig:rhoprpt}
\end{figure}

Upon using the ansatz for redshift and shape function
that yield the line element of 
Eq.~\eqref{metricconcrete}
and the stress-energy tensor in the form of
Eq.~\eqref{tabani} in the field equation of the quadratic gravity
theory we are considering, Eq.~\eqref{field}, one obtains three
equations, for $\rho=\rho(r)$, $p_r=p_r(r)$, and $p_t=p_t(r)$ in terms
of the functions $\zeta\left(r\right)$, $b\left(r\right)$, and their
first and second derivatives.
These equations in general are long and not
particularly useful, and thus we do not present them here.
As an interesting specific case,
we consider the throat to be at the radius $r_0=2M$, where
$M$ is a constant with mass units yet to be specified.
We obtain
\begin{equation}
\begin{aligned}
\rho(r)=&-\frac{M^2}{4\pi r^{12}}
\left[2r^8+160Mr^5\alpha\zeta_0+32M^5r\alpha\zeta_0^3+
+16M^6\alpha\zeta_0^4-24r^6\alpha\left(4+\zeta_0^2\right)
\right.
\\
&
\left.
-8M^3r^3\alpha\zeta_0
\left(100+\zeta_0^2\right)-8M^4r^2\alpha\zeta_0^2
\left(142+\zeta_0^2\right)+
M^2r^4\alpha\left(496+368\zeta_0^2+\zeta_0^4\right)\right]\,,
\quad r_0\leq r<r_\Sigma\,,
\\
p_r(r)=&\frac{M}{4\pi r^{12}}\left\{-r^9\zeta_0-96M^6r\alpha\zeta_0^3+
16M^7\alpha\zeta_0^4
-8M^5r^2\alpha\zeta_0^2\left(\zeta_0^2-30\right)
\right.
\\
&
+40M^4r^3\alpha\zeta_0\left(8+\zeta_0^2\right)
+M^3r^4\alpha
\left(\zeta_0^4-120\zeta_0^2-208\right)
-2Mr^6\left[r^2-8\alpha\left(4+\zeta_
0^2\right)\right]
\\
&
+\left.
4M^2r^5\zeta_0\left[r^2-\alpha\left(24+\zeta_0^2\right)\right
]\right\}\,,\hskip 7cm\quad r_0\leq r<r_\Sigma\,,
\\
\label{rhoprpt}
p_t(r)=&\frac{M}{8\pi r^{12}}\left\{r^9\zeta_0-512M^6r
\alpha\zeta_0^3-32M^7\alpha\zeta_0^4
+16M^5r^2\alpha\zeta_0^2\left(\zeta_0^2-190\right)
\right.
 \\
&
+
Mr^6\left(r^2-64\alpha\right)\left(4+\zeta_0^2\right)
+16M^4r^3\alpha\zeta_0\left(13\zeta_0^2-96\right)
 \\
&
\left.
-4M^2r^5\zeta_0\left[2r^2+5\alpha\left(\zeta_0^2-16\right)\right]
-2M^3r^4\left[2r^2\zeta_0^2+\alpha\left(\zeta_0^4
-488\zeta_0^2-624\right)
\right]\right\}\,,\quad\quad r_0\leq r<r_\Sigma\,,
\end{aligned}
\end{equation}
where the equations are valid for
$r_0\leq r\leq r_\Sigma$
in the
upper and lower domains of the interior
regions,  i.e.,  for  the two 
sides of the throat in the two domains of
the interior region up to the two mouths.
Different wormhole interior region
solutions can be
analyzed by
varying the parameter $\alpha$, which in turn 
yield different
quadratic theories, as well as by
varying the constants $\zeta_0$ and
$r_0$, which here we fixed to $r_0=2M$.
In Fig.~\ref{fig:rhoprpt} we provide plots of 
the energy density $\rho$, the radial pressure
$p_r$, and the tangential pressure $p_t$ as  functions of $r$,
for some specific
combination of values of
the parameters of the solution, namely,
$\alpha=0.013$ in Planck units, $\zeta_0=-60$, and
$r_0=2M$, for some parameter $M$.

\begin{figure}[h!]
\includegraphics[scale=0.7]{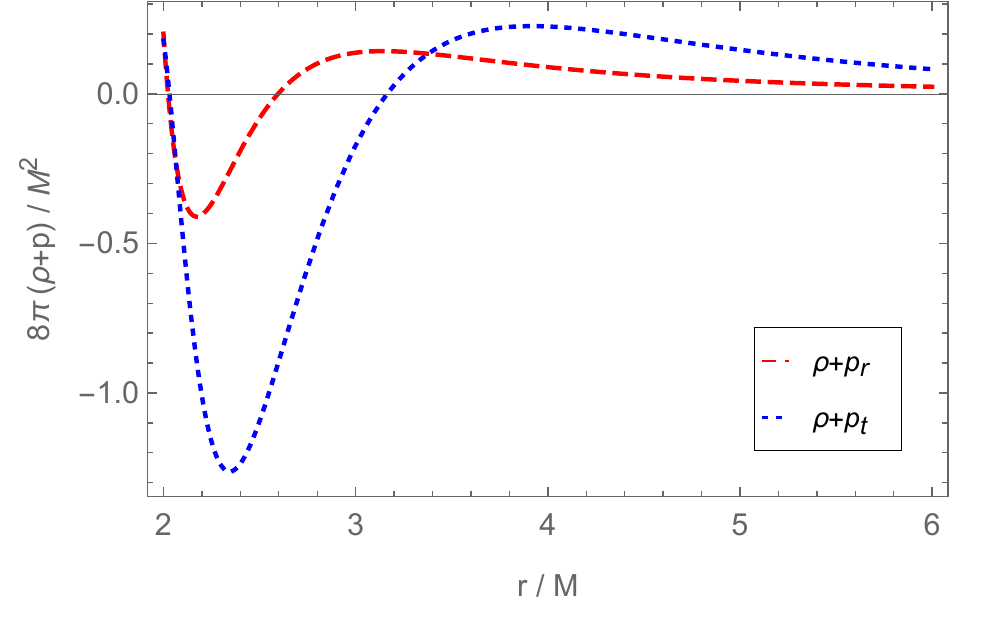}
\caption{
Plots of $\rho+p_r$ and $\rho+p_t$ as functions of the radius $r$ for
$\alpha=0.013$ in Planck units, $\zeta_0=-60$, and $r_0=2M$, where
here $M$ is some mass constant.  Clearly the NEC is satisfied at the
throat $r=r_0$ but violated elsewhere.  The quantities plotted are
normalized to $M$.}
\label{fig:nec}
\end{figure}

Having obtained the
solutions for the matter fields, one computes
$\rho+p_r$ and $\rho+p_t$
to verify whether the NEC is satisfied
at the throat $r_0$, which here is
$r_0=2M$, or not.
In Fig.~\ref{fig:nec} we provide plots of 
$\rho+p_r$ and $\rho+p_t$  as functions of $r$ for some specific
combination of values of
the parameters of the solution, namely, the same as
in Fig.~\ref{fig:rhoprpt}, i.e.,
$\alpha=0.013$ in Planck units, $\zeta_0=-60$, and
$r_0=2M$, for some parameter $M$.
One confirms
that the NEC is satisfied at the
wormhole throat $r=r_0=2M$, but it is ultimately
violated at
larger values of $r$.
To have an interior  solution
that satisfies the NEC throughout
the entire interior region one must truncate the solution
before the NEC starts to be violated.
From Fig.~\ref{fig:nec} the NEC starts to be violated
for radii $\frac{r}{M}$ outside the throat, although close
to it, so that for this set of parameters
one should put an end to the interior solution at about
$r_\Sigma=2.024M$.
We recall that
to have 
the matter NEC and the flaring out condition both
satisfied
at the throat
is only possible because we are dealing
with a modified gravity theory, here a
quadratic theory of gravity with an $\alpha R^2$
term. 
In this case, it is possible
to find a combination of
parameters
that fulfills the NEC at the throat as we showed,
noting also that there are many other
combinations of parameters 
for which the NEC is
violated there.
On the other
hand, in general relativity, in which case $\alpha=0$, the NEC is
always violated at the throat.
To see this, in the
Appendix~\ref{necgeneralrelativity}
we show that for $\alpha=0$,
the quantity $\rho+p_r$
of the solution presented, is negative, violating thus the NEC,
as expected.

\subsubsection{The exterior made of two symmetric domains}

We have now to provide an exterior solution to be matched with the
interior solution presented above, in order to form the whole wormhole
spacetime solution.  To find an exterior solution we have to specify
the redshift and shape functions, $\zeta\left(r\right)$ and $b(r)$,
respectively, that appear in Eq.~\eqref{metric}.  A specific form for
these functions is $\zeta\left(r\right)=\ln
\left(1-\frac{2M}{r}\right)+\zeta_e$ and $b\left(r\right)=2M$, where
$\zeta_e$, a constant to be determined later, and $M$ is a parameter
that represents the exterior spacetime mass.  With this choice the
line element of Eq.~\eqref{metric} suitable for the exterior region
made of two domains, the upper and lower, 
is
\begin{align}\label{schwarz}
ds^2=-\left(1-\frac{2M}{r}\right)e^{\zeta_e}dt^2+\left(1-
\frac{2M}{r}\right)^{-1}dr^2 + r^2d\Omega^2
\,,\quad\quad\quad\quad r_\Sigma<r<\infty\,, 
\end{align}
where we avoided to put a $+$ in all quantities, as it should be for
quantities in the $\mathcal V^+$ region, in order to not overload the
formulas with symbols.
Note that $\zeta_e$
ensures that the time
coordinate of both regions,
the interior wormhole and the exterior regions,
is the same upon matching.
The radial coordinate in Eq.~\eqref{schwarz}
ranges from
$r_\Sigma$ to infinity, $r_\Sigma< r <\infty$,
the upper domain of the exterior region starting
at the upper boundary $r_\Sigma$, the upper mouth, and
extending to infinite radial coordinate,
and 
the lower domain of the exterior region starting
at the lower boundary $r_\Sigma$, the lower mouth, and
extending to infinite radial coordinate.
Having the same form of line element on both sides
of the exterior region, the wormhole spacetime
has thus two symmetric domains in the exterior region.

The stress-energy tensor considered for the exterior $\mathcal V^+$,
constituted of the upper and lower domains, is that of an anisotropic
perfect fluid, which can be written as
\begin{equation}\label{tabexterior}
{T_a}^b(r)=\text{diag}\left(-\rho(r),p_r(r),p_t(r),p_t(r)\right)
\,,\quad\quad\quad\quad r_\Sigma<r<\infty\,,
\end{equation}
where again we avoided to put a $+$ in all quantities, as it should be
for quantities in the two domains of the exterior region $\mathcal
V^+$, in order to not overload the formulas.
The energy density
$\rho(r)$, the radial pressure $p_r(r)$, and the transverse pressure
$p_t(r)$, depend on the radial
coordinate alone, so that the matter distribution is static and
spherically symmetric.

Upon using the ansatz for redshift and shape function
that yield the line element of 
Eq.~\eqref{schwarz}
and the stress-energy tensor in the form of
Eq.~\eqref{tabexterior} in the field equation of the quadratic gravity
theory we are considering, Eq.~\eqref{field}, one obtains three
equations,
which in this case are simply
\begin{align}
\rho(r)=0\,,\quad
p_r(r)=0\,,\quad
p_t(r)=0\,,
\quad\quad\quad\quad r_\Sigma<r<\infty\,,
\label{rhoprptexterior}
\end{align}
where the equations are valid for
$r_\Sigma<r<\infty$
in the
upper and 
lower domains,  i.e.,  for  the two 
sides of the mouths in the exterior up to infinity.
Thus, the two domains of the exterior region
are vacua, i.e., with $T_{ab}=0$,
and so the solution for them
is the Schwarzschild solution, an asymptotically
flat solution. 
Indeed, inserting the vacuum assumption into Eq.~\eqref{field}, one
finds
that the modified field equations are satisfied for any
metric with $R_{ab}=0$, and consequently
with $R=0$.

Since the two domains of the exterior solution
are vacuum solutions, $T_{ab}=0$, they
satisfy the NEC, $T_{ab}k^ak^b\geq0$, automatically.  We have now
provided an exterior spacetime valid for the two
symmetric domains of the exterior region. This exterior
solution is to be
matched with the interior spacetime solution presented above.  In
order to form the whole wormhole spacetime solution, we need to apply
the junction conditions at the two junction
boundaries which we assume both to have radii $r_\Sigma$.

\subsubsection{Matching the wormhole interior to the vacuum exterior}
\label{sec:matches}

Before we apply the junction conditions,
Eqs.~\eqref{junction1}-\eqref{sab},
we need
to compute the relevant quantities on which they depend,
namely,
$h_{00}^-$,
$h_{00}^+$,
$n^{-c}$,
$n^{+c}$,
${{K^-}_{0}}^0$,
${{K^-}_{\theta}}^\theta$,
${K^-}$,
${{K^+}_{0}}^0$,
${{K^+}_{\theta}}^\theta$,
${K^+}$,
$R^{-\Sigma}$,
$R^{+\Sigma}$,
$(\nabla_cR)^{-\Sigma}$, and
$(\nabla_cR)^{+\Sigma}$,
for the wormhole spacetime regions we are considering.  These
expressions, which have to be computed from the interior line element
of Eq.~\eqref{metricconcrete} with its redshift and shape functions,
and from exterior
vacuum solution described by the line element
of Eq.~\eqref{schwarz},
are given in the Appendix~\ref{formulasanddefinitions}, where the
exponents $\gamma$ and $\beta$ for the interior metric functions are
put as generic,
whereas here we adapt those expressions to $\gamma=1$
and $\beta=1$.
In the junction conditions, Eqs.~\eqref{junction1}-\eqref{sab},
we set $\epsilon=1$ which is the value appropriate for
a timelike thin shell, the case we are considering.

The first junction condition, i.e., Eq.~\eqref{junction1}, dictates
the continuity of the metric through
each of the two boundaries $\Sigma$.  For
the two line elements
in Eqs.~\eqref{metricconcrete} and \eqref{schwarz} the
junction condition $\left[h_{ab}\right]=0$ yields
\begin{equation}\label{setzetae}
e^{\zeta_0\left(\frac{r_0}{r_\Sigma}\right)}=
\left(1-\frac{2M}{r_\Sigma}\right)e^{\zeta_e}\,,
\quad\quad\quad\quad\quad r=r_\Sigma\,,
\end{equation}
and thus, given $\zeta_0$ from the model chosen, one finds a relation
between $r_\Sigma$ and $\zeta_e$. Now, $r_\Sigma$ is defined as the
radius at which the matching is performed, and there is one $r_\Sigma$
for the matching of the upper domain and another $r_\Sigma$ for the
matching of the lower domain.

The second junction condition, i.e.,
Eq.~\eqref{junction2}, yields a constraint
on the radius $r_\Sigma$. For the
two line elements in
Eqs.~\eqref{metricconcrete} and \eqref{schwarz}, the junction
condition $\left[K\right]=0$ yields
\begin{equation}\label{setrsigma}
\frac{2r_\Sigma-3M}{r_\Sigma^2\sqrt{1-\frac{2M}{r_\Sigma}}}=
\frac{1}{2r_\Sigma}\sqrt{1-
\left(\frac{r_0}{r_\Sigma}\right)^{2}}
\left[4-\zeta_0\left(\frac{r_0}{r_\Sigma}\right) \right]\,,
\quad\quad\quad\quad\quad r=r_\Sigma\,.
\end{equation}
For each chosen $\zeta_0$, one can find $r_\Sigma$ from
Eq.~\eqref{setrsigma}. The matching between the two metrics must then
be performed at the $r_\Sigma$ found, one $r_\Sigma$ for the upper
domain and another for the lower domain.  One can then insert into
Eq.~\eqref{setzetae} the value of $r_\Sigma$ found, and obtain the
corresponding value of $\zeta_e$.

The third junction condition, i.e., Eq.~\eqref{stressts}, is now used.
We assume that each thin shell, one in the upper
domain, the other in the lower domain, is made of an isotropic perfect fluid.
For a perfect fluid, the surface energy density
$\sigma$ and the tangential pressure $p$ of each thin shell
are found from the surface stress-energy tensor
${S_{\alpha}}^\beta=S_{ac}e_\alpha^ae^c_\gamma  h^{\beta\gamma }$,
where, as defined before,
the vectors $e^a_\alpha$
represent the projection vectors
that project from the four-dimensional
spacetime region  to the hypersurface $\Sigma$. The tensor 
${S_\alpha}^\beta$
takes
a diagonal form, and we can write
$
{S_\alpha}^\beta=\text{diag}\left(-\sigma,p,p\right)
$.
Defining the surface stress-energy tensor
with full spacetime indices as $S_{ab}=
e_a^\alpha e_b^\beta S_{\alpha\beta}$, the third
junction condition of the theory, namely, 
$\left[K_{ab}\right]+
2\alpha\left(h_{ab}n^c\left[\nabla_cR\right]-
R^\Sigma\left[K_{ab}\right]-K_{ab}^\Sigma\left[R\right]\right)
=8\pi S_{ab}$, 
yields 
the surface energy density $\sigma$ and
the transverse pressure $p$ of each thin shell as
having the expressions 
$\sigma = \frac{1}{8\pi}
\Big[
\left[K_0^0\right]
\left(1+2\alpha R_\Sigma\right)+
2\alpha\left(K^{0(\Sigma)}_{0}\left[R\right]-n^c
\left[\nabla_c R\right]\right)\Big]$, and
$p=\frac{1}{8\pi}
\Big[
\frac{1}{2}
\left[K_0^0\right]
\left(1+2\alpha R_\Sigma\right)
+
\alpha
\left(K^{0(\Sigma)}_{0}
\left[R\right]+
2n^c\left[\nabla_c R\right]\right)
\Big]
$,
where we
used the fact that, since $\left[K\right]=0$ holds, then
in spherically symmetric spacetimes and spherical coordinates we have
$\left[K_0^0\right]=-
2\left[K_\theta^\theta\right]=-2\left[K_\phi^\phi\right]$.
One can thus write $\sigma$ and $p$
for each of the two thin shells,
one in the upper domain and the other in the lower domain,
as
\begin{equation}
\begin{aligned}
\label{press}
\sigma = &\frac{1}{8\pi}
\Bigg[ \,\,a(r_\Sigma)\,
\Big(1\,+\,2\alpha f(r_\Sigma)\Big)+
2\alpha\Big( b(r_\Sigma)g(r_\Sigma)-h(r_\Sigma)\Big)
\Bigg]\,,\quad\quad\quad\quad\quad r=r_\Sigma\,,
\\
p = &\frac{1}{8\pi}
\Bigg[ \frac12 a(r_\Sigma)
\Big(1+2\alpha f(r_\Sigma)\Big)+
\alpha\Big( b(r_\Sigma)g(r_\Sigma)+2h(r_\Sigma)\Big)
\Bigg]\,,
\quad\quad\quad\quad\quad r=r_\Sigma\,,
\end{aligned}
\end{equation}
where 
$a(r_\Sigma)$,
$b(r_\Sigma)$,
$f(r_\Sigma)$,
$g(r_\Sigma)$, and
$h(r_\Sigma)$,
are functions of
$r_\Sigma$ and
depend also on the remaining free parameters that characterize
the interior and exteriors regions, see the 
Appendix~\ref{formulasanddefinitions} for the explicit
expressions.
For the surface
stress-energy tensor $S_{ab}$, part of the
full stress-energy tensor $T_{ab}$ of Eq.~\eqref{tabquad},
one has to verify whether the NEC is satisfied or not. 
Since 
one
can write $S_{ab}$ as $S_{ab}=
e_a^\alpha e_b^\beta S_{\alpha\beta}$,
and 
${S_\alpha}^\beta=\text{diag}\left(-\sigma,p,p\right)$,
one has 
in an
orthonormal tetrad that $S_{ab}=(\sigma,0, p, p)$. 
For a null
vector at $\Sigma$ of the form $k^a=(1,0,1,0)$ or of the form
$k^a=(1,0,0,1)$, i.e., null vectors on the shell, one has
$S_{ab}k^ak^b= \sigma+ p$, so that for vectors on the shell $\Sigma$,
the NEC is satisfied if $\sigma+p\geq 0$.  For a null vector at $\Sigma$
of the form $k^a=(1,1,0,0)$ one has $S_{ab}k^ak^b=\sigma$ and for this
type of vectors, i.e., null vectors off the shell,
the NEC implies $\sigma\geq 0$.  The three types of
null vectors are representative of all null vectors at $\Sigma$, so to
satisfy the NEC, the fluid quantities must satisfy $\sigma+p\geq 0$ and
$\sigma\geq 0$, which from Eq.~\eqref{press} are conditions
that depend on the functions $a(r_\Sigma)$,
$b(r_\Sigma)$, $f(r_\Sigma)$, $g(r_\Sigma)$, and $h(r_\Sigma)$.

The fourth junction condition, given
in 
Eq.~\eqref{s}, i.e.,
$2\alpha K^\Sigma\left[R\right]=8\pi P$, yields
for each of the two thin shells,
one in the upper domain and the other in the lower domain,
the following result
\begin{equation}\label{snew}
 P=\frac{\alpha}{4\pi}
 c(r_\Sigma)
g(r_\Sigma)\,,
\quad\quad\quad\quad\quad r=r_\Sigma\,,
\end{equation}
where $c(r_\Sigma)$ and $g(r_\Sigma)$
are functions of
$r_\Sigma$ and
also depend on the remaining free parameters
that characterize
the interior and exteriors regions, see the 
Appendix~\ref{formulasanddefinitions} for the explicit
expressions.
$P$ 
measures the external normal pressure
supported on each thin shell, the upper
and the lower ones.
In the stress-energy tensor given in Eq.~\eqref{tabquad}
the term that corresponds to this junction condition
is $Pn_an_b$,
with the normal
being $n_{-a}=(0,
\left(
1-\left(\frac{r_0}{r_\Sigma}
\right)^{2}
\right)^{-\frac12}
,0,0)$ and
$n_{+a}=(0,
\left(
1-\frac{2M}{r_\Sigma}
\right)^{-\frac12}
,0,0)$,
see the 
Appendix~\ref{formulasanddefinitions}.
The normals  $n_{-a}$ and  $n_{+a}$
are the same at $r_\Sigma$, only
written in different coordinate systems,
indeed $n_{\mp a}
n^{\mp a}=1$ and they are applied at the same point.
Since this part of
the stress-energy tensor at the thin shell
has the form 
$P\,n_an_b$,
the NEC for it can be written as 
$P\,n_an_bk^ak^b=P\,|n_ak^a|^2\geq0$, which is equivalent to 
$P\geq0$, which in turn
depends on the functions
$c(r_\Sigma)$ and 
$g(r_\Sigma)$.

The fifth junction condition, given in 
Eq.~\eqref{sa}, i.e.,
$-2 \alpha h^c_a\nabla_c\left[R\right]=8\pi F_a$,
involves $F_a$, which is the external flux momentum,
the time component of which
determines the energy flux across $\Sigma$,
while
the spatial components determine the
corresponding tangential stresses.
Since the solution is spherically symmetric,
one has that 
$\nabla_c\left[R\right]$ is a function of $r$
only, 
and since 
the induced metric projects
tensors into a hypersurface of constant radius, the contraction
$h_a^c\nabla_c\left[R\right]$ in Eq.~\eqref{sa} vanishes,
$F_a=0$. More specifically,
although $\nabla_c\left[R\right]$ is
nonzero, the only nonvanishing component of this tensor is
along the radial direction and so its projection
with $h^c_a$
into the hypersurface vanishes. So,
\begin{equation}\label{sanew}
F_\alpha=0\,,
\quad\quad\quad\quad\quad r=r_\Sigma
\,.
\end{equation}
Thus, each of the two thin shells, the upper and
the lower, support zero external flux momentum.
In the stress-energy tensor given in Eq.~\eqref{tabquad},
the term that corresponds to this junction condition
is $F_{(a}n_{b)}$.
It satisfies the NEC if
$F_{(a}n_{b)}k^ak^b\geq0$, which, since $F_a=0$,  is the case
without further specification.

The sixth junction condition, given in Eq.~\eqref{sab},
i.e., $\frac{\alpha}{4\pi}
h_{ab}[R]n^c\nabla_c\delta\left(l\right)=s_{ab}\left(l\right)$,
has to be worked out with care.
Since $h_{ab}$ is a diagonal tensor, we can write
the double layer thin shell stress-energy tensor 
$s_{ab}\left(l\right)$ as 
${s_a}^b\left(l\right)=\text{diag}
\left(-\bar\sigma\left(l\right),0,\bar p\left(l\right),\bar
p\left(l\right)\right)$,
where $\bar\sigma\left(l\right)$ and $\bar
p\left(l\right)$ are defined to be
the double layer surface density and tangential
pressure distributions, respectively.
Then, for this wormhole Eq.~\eqref{sab} gives
\begin{align}\label{6thgenexample}
\bar\sigma=-\frac{\alpha}{4\pi}g(r_\Sigma)\,,\quad\quad
\bar p=\frac{\alpha}{4\pi}g(r_\Sigma)
\,,
\quad\quad\quad\quad\quad r=r_\Sigma\,,
\end{align}
where we have defined $\bar\sigma$ and $\bar p$ as the corresponding
quantities without $n^c\nabla_c\delta\left(l\right)$, i.e., $\bar
\sigma\left(l\right)=\bar \sigma\, n^c\nabla_c\delta\left(l\right)$
and $\bar p\left(l\right)=\bar p \,n^c\nabla_c\delta\left(l\right)$,
and where $g(r_\Sigma)$ is a function of $r_\Sigma$ and also depends
on the remaining free parameters that characterize the interior and
exteriors regions, see the Appendix~\ref{formulasanddefinitions} for
the explicit expression.  These are the expressions for the double
layer surface energy density and surface pressure that each thin
shell, one in the upper domain, another in the lower domain, must
satisfy.  We have now to check the NEC for this double layer part of the
stress-energy tensor, $s_{ab}(l)$. Since $s_{ab}(l)$ is
proportional to a
derivative of a delta function, namely,
$n^c\nabla_c\delta\left(l\right)$, it is not enough to verify whether
$s_{ab}(l)k^ak^b\geq0$ is satisfied at the thin shell. One has to verify
the inequality in the neighborhood of  the thin shell, i.e., whether
$\int s_{ab}(l)k^a k^b dl\geq0$ holds, for $k^a$ any null vector.  A
null vector $k^a$ satisfies $g_{ab}k^ak^b=0$, and so it can be written,
apart some conformal factor, as
$k^a=\left(\frac{-1}{\sqrt{g_{tt}}},\frac{a}{\sqrt{g_{rr}}},
\frac{b}{\sqrt{g_{\theta\theta}}},
\frac{c}{\sqrt{g_{\phi\phi}}}\right)$, where the parameters $a$, $b$
and $c$ satisfy the
condition $a^2+b^2+c^2=1$.
One can now compute $\int s_{ab}(l)k^a k^b dl$.
Putting $s\equiv\frac{\alpha}{4\pi}g(r_\Sigma)$ to shorten the
notation, one has $s_{ab}\left(l\right)= s
\,h_{ab}n^c\nabla_c\delta\left(l\right)$.  Thus, $\int s_{ab}(l)k^a
k^b dl =\int s \,h_{ab}k^a k^b n^c\nabla_c\delta\left(l\right)dl$.
One can perform this integral by parts to obtain $ \int s \,h_{ab}k^a
k^b n^c\nabla_c\delta\left(l\right)dl = \int n^c\partial_c\left[ s
\,h_{ab}k^a k^b \delta\left(l\right)\right]dl -\int n^c\nabla_c
\left(s \,h_{ab} k^a k^b \right) \delta\left(l\right) dl $.  The first
term,
$\int n^c\partial_c\left[ s
\,h_{ab}k^a k^b \delta\left(l\right)\right]dl$,
can be integrated directly and can be seen
to
vanish because at the limits of integration in $l$ the
function $\delta(l)$ is zero, i.e., $\left[ s \,h_{ab}k^a k^b
\delta\left(l\right)\right]_{l_{\rm min}}^{l_{\rm max}}=0$.  The
second term,  $-\int n^c\nabla_c
\left(s \,h_{ab} k^a k^b \right) \delta\left(l\right) dl$,
however, does not vanish.  Indeed, since $s$ is a
constant  
one can write $-n^c\nabla_c \left(s \,h_{ab} k^a k^b \right) =-s
n^c\nabla_c \left( h_{ab} k^a k^b \right)$.  Now, on any neighborhood
of $\Sigma$, the null vector $k^a$ can be any, and so the term
$h_{ab}k^ak^b$ can be different along any radial path in several
manners, in certain radial paths it is negative, in others it is zero,
and yet in some others is positive, and thus the integral
can have negative
values for arbitrary $k^a$. So, the NEC condition on the
stress-energy component $s_{ab}(l)$ is not satisfied.
Moreover, even at $\Sigma$, the NEC may not be satisfied.
In this case one needs to check that
$s_{ab}k^ak^b\geq0$ at $\Sigma$.
As we have seen above for $S_{ab}$, now repeated for
$s_{ab}$, this means 
$\bar\sigma\geq0$ and $\bar\sigma+\bar p\geq0$.
Here, from Eq.~\eqref{6thgenexample}
one has $\bar\sigma+\bar p=0$  and 
the sign of $\bar\sigma$  depends
 on the function $g(r_\Sigma)$.

\subsection{Full solution for a
traversable wormhole with matter and double layer
thin shell}

We now provide specific values for the parameters appearing in this
wormhole solution in order to find an explicit solution.

Let us choose a quadratic theory of gravity
for which the coupling constant $\alpha$ has the value
\begin{equation}
\alpha=0.013 
\,,
\end{equation}
in Planck units. In fact, one should write $\alpha=0.013
\left(\frac{1}{M_{\rm pl}}\right)^2$ where $M_{\rm pl}$ is the Planck
mass, but in the natural system of units considered here, the Planck
mass is one, and thus $\alpha=0.013$.  We take the exterior mass $M$
as the scale for the whole solution. We can now describe a full
solution composed of the two symmetric domains of the interior region,
the two double layer thin shells at the mouths $\Sigma$, and the two
symmetric asymptotically flat domains of the exterior region.

For the two symmetric
domains of the interior, we have the line element Eq.~\eqref{metricconcrete},
and choose for  redshift constant $\zeta_0$ the value
\begin{equation}
\zeta_0=-60\,,
\end{equation}
and  for the throat
radius $r_0$  the value
\begin{equation}
r_0=2M\,.
\end{equation}
These are the same numbers as chosen for the solution plotted in
Figs.~\ref{fig:rhoprpt} and \ref{fig:nec}.  We have seen that for
these parameters the NEC is satisfied at the throat of the wormhole
$r_0=2M$, and is indeed satisfied up to the radius $r_\Sigma$.  We can
also note in passing that, for these parameters, the WEC is also
satisfied at the throat since $\rho>0$, although this feature per se
is not of major physical importance, since the WEC is violated
elsewhere.

For the two thin
shells, one at each mouth,  the results are the
following.
The first junction condition, Eq.~\eqref{setzetae},
gives 
\begin{equation}
\zeta_e=-54.847\,.
\end{equation}
The 
second junction condition, Eq.~\eqref{setrsigma}, 
gives 
\begin{equation}
r_\Sigma=2.024M\,.
\end{equation}
The third junction condition, Eq.~\eqref{press},
gives
\begin{equation}
\sigma=-\frac{1}{8\pi}
\frac{58.642}{M}\,,\quad\quad
p=\frac{1}{8\pi}\frac{62.964}{M}\,,
\label{sigmap}
\end{equation}
which are the values of the energy density and the
pressure at each of the thin shells, the upper
one and the lower one.
From Eq.~\eqref{sigmap}
we have
$(\sigma+p)=\frac{1}{8\pi}
\frac{4.322}{M}$.
Thus, although $\sigma+p>0$ and thus the NEC is satisfied
for null vectors on the shell, 
one has $\sigma<0$, meaning that the NEC is not satisfied
for null vectors out of the shell, 
and so this piece of
the thin shell stress-energy tensor 
does not satisfy  the NEC.
The fourth junction condition, Eq.~\eqref{snew},
gives the external normal
pressure at each thin shell,
\begin{equation}
P=\frac{1}{8\pi}\dfrac{1.515}{M}
\,.
\end{equation}
Since $P>0$, it implies that this piece of
the thin shell stress-energy tensor 
satisfies the NEC.
The fifth junction condition, Eq.~\eqref{sanew},
gives the external 
flux momentum at each thin shell, which vanishes,
\begin{equation}
F_\alpha=0
\,.
\end{equation}
Since $F_\alpha=0$, it implies that this piece of
the thin shell stress-energy tensor 
satisfies the NEC.
The sixth junction condition, Eq.~\eqref{6thgenexample},
gives
the stress-energy tensor distribution of the corresponding
double layer at each of the
two thin shells,
which for our particular choice of parameters yields
\begin{align}\label{sabnew}
\bar\sigma=-\frac{1}{4\pi}
\frac{
0.319}{M^2}\,,\quad\quad
\bar
p=\frac{1}{4\pi}
\frac{
0.319}{M^2}
\,.
\end{align}
Thus, $\bar\sigma+ \bar p=0$
and 
$\bar\sigma<0$.
It implies that this piece of
the thin shell stress-energy tensor 
does not satisfy the NEC.

In summary, this traversable wormhole is a symmetric wormhole
solution. It presents combinations of $\rho+p_r$ and $\rho+p_t$ for
the two domains of the interior region that are positive in the range
of $r$ between the wormhole throat radius $r_0$ and the shell radius
$r_\Sigma$.  At the the two shell radii, $r_\Sigma$, the solution
yields $\sigma+p>0$ with $\sigma<0$, $P>0$, $F_\alpha=0$, and
$\bar\sigma+ \bar p=0$ with $\bar\sigma<0$. For the exterior region,
$r>r_\Sigma$, the stress-energy tensor vanishes and so the NEC is
satisfied. Thus, we see that the stress-energy tensor has
two components
that do not satisfy the NEC.
Although we have not proved, in practical
terms we found it impossible
to have this type of traversable wormholes
with thin shell double layers in quadratic gravity
having the NEC satisfied for all pieces of the
wormhole matter stress-energy
tensor.

%%%%%%%%%%%%%%%%%%%%%%%%%%%%%%%%%%%%%%%%%%%%%%%%%%%%%%%%%%%%%%%%
\section{Thin shell traversable wormholes with a single layer in
quadratic $R+\alpha R^2$ gravity}
\label{sec:wormssinglelayer}
%%%%%%%%%%%%%%%%%%%%%%%%%%%%%%%%%%%%%%%%%%%%%%%%%%%%%%%%%%%%%%%%

\subsection{General considerations}

A thin shell traversable wormhole is defined as a traversable wormhole
with no interior and with the thin shell located precisely at the
throat.  Here, we construct a thin shell traversable wormhole with a
single layer. For the construction of a thin shell traversable
wormhole with a double layer see \cite{eiroa1}.

In a thin shell traversable wormhole, the two domains of the
exterior region are matched directly to each other at a given
throat radius $r_0$, which
in turn coincides with the matching boundary
radius $r_\Sigma$, i.e., the throat and the two mouths coincide. Since
we are interested in a wormhole solution for which the matching is
done with a single layer thin shell at the separation boundary, the
double layer junction conditions, Eqs.~\eqref{junction1}-\eqref{sab},
can be used noting that they reduce by a careful examination to a
single layer thin shell matching, as shown in Sec.~\ref{sec:jcs}.  The
wormhole to be constructed is asymmetric, i.e., the two domains of the
exterior region, situated at each side of the throat, are different.
${\mathcal V}^+$ is now the upper domain, say, and ${\mathcal V}^-$ is
now the lower domain, the whole wormhole spacetime ${\mathcal V}$ is
composed of ${\mathcal V}^+$ and ${\mathcal V}^-$ plus the thin shell
at the throat $\Sigma$.

\subsection{Asymmetric single layer  thin shell
traversable wormhole solutions satisfying the NEC and all
the energy conditions at the throat}
\label{sec:wormsol0}

\subsubsection{The interior}

By definition,
the interior of a thin shell traversable wormhole
is an empty set, i.e., it has no line
element and no stress-energy tensor, so it is degenerated.

\subsubsection{The exterior made of two asymmetric domains}

For the exterior region made of two domains,
the upper and the lower, one
possibility
would be to consider two
vacuum Schwarzschild spacetimes with
the same mass $M$. However, this
is not feasible, since the third junction condition 
$\left[K\right]=0$ would not
be satisfied at any radius $r_\Sigma$. We thus consider
instead two Schwarzschild spacetimes, the
upper exterior domain with positive mass,
$M$, and the lower exterior domain with negative
mass, $-M$, thus it is a domain
of gravitational repulsion.
In this case the line elements for the
two  spacetime domains of the exterior region,
the upper and lower
domains, are
\begin{equation}
\begin{aligned}
ds^2=&-\left(1-\frac{2M}{r}\right)e^{\zeta_u}dt^2+
\left(1-\frac{2M}{r}\right)^
{-1}dr^2+r^2d\Omega^2\,,\quad\quad
r_0=r_\Sigma\leq r<\infty\quad{\rm upper \; domain}\,,
\label{positivemass}
\\
ds^2=&-\left(1+\frac{2M}{r}\right)\;\;dt^2\;+\,
\left(1+\frac{2M}{r}\right)^
{-1}dr^2+r^2d\Omega^2\,,\quad\quad
r_0=r_\Sigma\leq r<\infty\quad{\rm lower \; domain}\,,
%\label{negmass}
\end{aligned}
\end{equation}
where the radial coordinate ranges from
$r_0=r_\Sigma$ to infinity, i.e., 
$r_0=r_\Sigma\leq r<\infty$,
for both
spacetime domains of the exterior region,
the throat radius $r_0$
and the two mouths at $r_\Sigma$ coincide in this
case, 
and $\zeta_u$ is a constant introduced
in the upper domain line element for later convenience.
Since the two
line elements
of Eq.~\eqref{positivemass} 
have $R_{ab}=0$, and are vacuum regions, i.e., $T_{ab}=0$, they
are solutions of
the quadratic gravity
field equation
given in Eq.~\eqref{field}.
These two exterior
spacetimes are to be matched at
the radius
$r_\Sigma$ with a single layer
thin shell.

\subsubsection{Matching the two domains of the
wormhole exterior: A single layer thin shell}

Let us now apply the junction conditions provided in
Eqs.~\eqref{junction1}-\eqref{sab} in the single layer matching
situation, see Sec.~\ref{sec:jcs}, to the two domains
of the wormhole exterior, 
Eq.~\eqref{positivemass}.
In the junction conditions
we set $\epsilon=1$ which is the value appropriate for
a timelike thin shell, the case we are considering.

The first junction condition, i.e., Eq.~\eqref{junction1}, 
$\left[h_{ab}\right]=0$ provides a constraint on the constant
$\zeta_u$ of the form
\begin{equation}\label{puretsjc1}
r_\Sigma+2M=\left(r_\Sigma-2M\right)e^{\zeta_u}
\,,
\end{equation}
where the radius $r_\Sigma$ at which the matching is
performed will be set by
the second junction condition.

The second junction condition, i.e., Eq.~\eqref{junction2},
$\left[K\right]=0$,
can be
written as
\begin{equation}\label{222}
\frac{2r_\Sigma-3M}{\sqrt{1-
\frac{2M}{r_\Sigma}}}=\frac{2r_\Sigma+3M}{\sqrt{1+\
\frac{2M}{r_\Sigma}}}.
\end{equation}
This equation can now be solved for $r_\Sigma$.
Then, the value found for
$r_\Sigma$
can be inserted into
Eq.~\eqref{puretsjc1} to obtain
the value of $\zeta_u$.

The third junction condition, i.e., Eq.~\eqref{stressts},
for a single layer matching is reduced to
the following form 
$-\left[K_{ab}\right]+
2\alpha\left(h_{ab}n^c\left[\nabla_c R\right]-
R^\Sigma\left[K_{ab}\right]\right)=8\pi S_{ab}$.
We assume that the thin shell is made of an isotropic perfect fluid.
For a perfect fluid,
the tensor 
${S_\alpha}^\beta$
takes
a diagonal form, and we can write
$
{S_\alpha}^\beta=\text{diag}\left(-\sigma,p,p\right)
$.
Defining the four-dimensional surface stress-energy tensor as $S_{ab}=
e_a^\alpha e_b^\beta S_{\alpha\beta}$, the third
junction condition for a single layer thin shell
yields 
the surface energy density $\sigma$ and
the transverse pressure $p$ of the thin shell as
having the expressions 
$\sigma=
\frac{\left[K_0^0\right]}{8\pi}$ and
$p=\frac{\left[K_0^0\right]}{16\pi}$.
Since $\left[K_0^0\right]=
\frac{M}{r_\Sigma^2}\left(\sqrt{\frac{r_\Sigma}{r_\Sigma-2M}}+
\sqrt{\frac{r_\Sigma}{r_\Sigma+2M}}  \right)$,
the full expressions are
\begin{align}
\sigma = \frac{1}{8\pi}
\frac{M}{r_\Sigma^2}\left(\sqrt{\frac{r_\Sigma}{r_\Sigma-2M}}+
\sqrt{\frac{r_\Sigma}{r_\Sigma+2M}}  \right)
\,,\quad\quad
p =
\frac{1}{8\pi}
\frac{M}{2r_\Sigma^2}\left(\sqrt{\frac{r_\Sigma}{r_\Sigma-2M}}+
\sqrt{\frac{r_\Sigma}{r_\Sigma+2M}}  \right)\,.
\label{presssingle}
\end{align}
Then, the value found for
$r_\Sigma$ in Eq.~\eqref{222}
can be inserted into
Eq.~\eqref{presssingle} to obtain
the values of $\sigma$ and $p$.
The NEC for $S_{ab}$ is satisfied if $S_{ab}k^ak^b\geq0$
for any null vector $k^a$.
In an
orthonormal tetrad one has $S_{ab}=(\sigma,0, p, p)$,
null vectors on the shell have the form 
$k^a=(1,0,1,0)$ or 
$k^a=(1,0,0,1)$,
and null vectors out of the shell have the form $k^a=(1,1,0,0)$.
This implies that the NEC is satisfied if $\sigma+p\geq 0$ and
$\sigma\geq 0$.

The fourth junction condition,
Eq.~\eqref{s}, which is now reduced to
$[R]=0$,
is automatically satisfied for any matching
radius $r_\Sigma$ since the two spacetimes given in
Eq.~\eqref{positivemass}
have $R_{ab}=0$ and so $R=0$.

The fifth and sixth junction conditions, Eqs.~\eqref{sa}
and~\eqref{sab}, respectively, are trivially satisfied, as
both sides of the equations vanish.

\subsection{Full solution for the single layer  thin shell
traversable wormhole}

For the thin shell traversable 
wormhole solution considered there are no free
parameters. The two
line elements given in Eq.~\eqref{positivemass}
depend
solely on the mass $M$ which is used as a scale in the problem, and
the coupling constant $\alpha$ does not play a role since both
domains of the exterior
region have $R=0$.  Thus, these thin shell wormholes with a single
layer and vacuum domains in the exterior region are solutions of
quadratic gravity for any value of $\alpha$, including $\alpha=0$,
i.e., general relativity. Note that the reverse
is not true, wormhole solutions in general relativity are
not by rule wormholes in quadratic theories of gravitation,
as the case we mentioned in the passing above is an example of,
i.e., two Schwarzschild solutions with the same positive mass $M$
matched at the throat
are solutions in general relativity but not solutions in
the quadratic theory presented here.
Let us describe in full the
wormhole.

The interior is an empty set, indeed there is no interior per se.

The matching thin shell is at a radius $r_\Sigma$
which coincides with the radius $r_0$
of the throat, $r_0=r_\Sigma$.
The  thin shell properties are determined
from the junction conditions.
The first junction condition, i.e., Eq.~\eqref{puretsjc1}, 
provides a constraint on the constant
$\zeta_u$,
where the radius $r_\Sigma$ at which the matching is
performed is set by
the second junction condition, i.e., Eq.~\eqref{222},
which yields
$r_\Sigma=\frac{3}{\sqrt{2}}\,M$.
This value
can be inserted into
Eq.~\eqref{puretsjc1} to obtain
$\zeta_u=\ln\left(17+12\sqrt{2}\right)$.
Thus, we have 
\begin{equation}\label{zetanow}
e^{\zeta_u}=17+12\sqrt{2}
\,,
\end{equation}
which is approximately $e^{\zeta_u}=33.981$,
and 
\begin{equation}\label{rSigmanow}
r_\Sigma=\frac{3}{\sqrt{2}}\,M\,,
\end{equation}
which is approximately $r_\Sigma=2.121$.
The third junction condition, i.e., Eq.~\eqref{presssingle},
then gives
\begin{align}\label{sigmapnow}
\sigma = 
\frac{1}{8\pi}\frac{1.089}{M}\,,\quad\quad
p = \frac{1}{8\pi}\frac{0.544}{M}\,.
\end{align}
Then, $\sigma+p=\frac{1}{8\pi}\frac{1.633}{M}$.
Since $\sigma+p$ and $\sigma$
are
positive, the NEC for the matter at the thin shell
is satisfied. Furthermore,
since also $p>0$, and $\sigma=2p$, one can show
that this actually
implies that all energy conditions are satisfied by the
matter at the thin shell, as
$\sigma\geq0$, $\sigma+p\geq0$, $\sigma+2p\geq0$, and
$\sigma\geq|p|$. Given that both the upper and lower spacetime
domains are
vacuum solutions, which automatically satisfy all energy conditions,
the single layer thin shell traversable wormhole solution obtained
satisfies all energy
conditions for the entire spacetime.  The fourth junction condition is
automatically satisfied. Note also that since
$r_\Sigma=\frac{3}{\sqrt{2}}\,M$
is approximately $r_\Sigma =2.121M$ one has that
$r_\Sigma$ satisfies $r_\Sigma>2M$.  Since the upper spacetime
domain solution given in 
Eq.~\eqref{positivemass}
has positive mass $M$ with $r_\Sigma>2M$, there is
no horizon on this side of the solution, and since the lower domain spacetime
solution given in  Eq.~\eqref{positivemass}
has negative mass $-M$, there is no
horizon from that side also, thus guaranteeing the traversability of
the solution.

In summary, this solution represents an asymmetric
traversable wormhole with a single layer thin shell
that
satisfies all the energy conditions. More specifically,
the wormhole has no interior and
possesses a single layer thin shell at the throat that joins an upper
domain, belonging to the exterior region, described by a
Schwarzschild spacetime with positive
mass to a lower domain, also belonging to the exterior region,
described by a Schwarzschild spacetime
with negative mass. In this wormhole,
whatever matter is attracted from the upper region into
the throat is expelled repulsively from the throat
into the lower region.

%%%%%%%%%%%%%%%%%%%%%%%%%%%%%%%%%%%%%%%%%%%%%%%%%%%%%%%%%%%%%%
\section{
Conclusions}
\label{sec:concl}
%%%%%%%%%%%%%%%%%%%%%%%%%%%%%%%%%%%%%%%%%%%%%%%%%%%%%%%%%%%%%

We have found asymptotically flat traversable wormhole
solutions 
in the 
quadratic theory of gravity
given by the following $f\left(R\right)$ function, 
$f\left(R\right)=R+\alpha R^2$.  One set of wormhole solutions has
double layer thin shells at the interfaces between an interior
with two domains made of nonexotic perfect fluid material
and a symmetric
exterior with two Schwarzschild spacetime domains.  These double
layers are characteristic of quadratic theories of gravity, and
although there are no two different types of charge in gravitation
they are analogous to dipole double layers in electrodynamics.
The NEC for these double layer traversable wormholes
has been tested
and it was found that
at the throat and for the whole
wormhole interior it is satisfied,
and at the two thin shells some stress-energy tensor
components satisfy it, while others, for instance, the double layer
distribution component, do not. 
Another set of wormhole solutions found has a single layer thin shell
at the the throat, characteristic
of thin shell traversable wormhole solutions,
and two asymmetric exterior Schwarzschild spacetimes,
one with positive mass, the other with
negative mass. The matter at the thin shell satisfies the NEC.

The set of six junction conditions in quadratic gravity for a double
layer matching implies fewer restrictions to the solutions, and thus
the method outlined to obtain traversable wormhole spacetimes with
double layer thin shells allows to find numerous solutions for a great
variety of combinations of parameters, although it seems that the NEC
is basically impossible, or at least very hard, to be satisfied when
double layer thin shells are present, at least in this quadratic
theory of gravity.  The set of four junction
conditions for a single layer matching, which can be found directly
from the double layer set of junction conditions, is more constrained,
and thus hampers the construction of traversable wormhole spacetimes,
although traversable thin shell wormholes, asymmetric ones, that
satisfy
the NEC at the throat were possible to construct. The set of three
junction conditions for a boundary surface matching, which can be
found directly from the previous sets of junction conditions,
restricts even further the constructions of this type of wormholes,
and we have not been able to find a solution with a smooth boundary
surface matching.

\acknowledgments{
We thank Kirill Bronnikov and Sergey Sushkov for conversations.  JLR
acknowledges the European Regional Development Fund and the programme
Mobilitas Pluss for financial support through Project No.~MOBJD647,
and project No.~2021/43/P/ST2/02141 cofunded by the Polish National
Science Centre and the European Union Framework Programme for Research
and Innovation Horizon 2020 under the Marie Sklodowska-Curie grant
agreement No. 94533.  RA acknowledges Funda\c{c}\~{a}o para a
Ci\^{e}ncia e Tecnologia - FCT, Portugal, for financial support
through project QEntHEP - Quantum Entanglement in High Energy Physics
EXPL/FIS-PAR/1604/2021.  RA and JPSL acknowledge Funda\c{c}\~{a}o para
a Ci\^{e}ncia e Tecnologia - FCT, Portugal, for financial support
through project UIDB/50008/2020.
}

\appendix

\section{The sixth junction condition
of the quadratic $R+\alpha R^2$ theory: the
double layer stress-energy tensor $s_{ab}\left(l\right)$}

\label{app0}

The sixth junction condition for the
thin shell of the quadratic $R+\alpha R^2$ theory
is given by the formula
\begin{equation}
\label{6thgeneralabapp0}
2\epsilon\alpha \, \Omega_{ab}\left(l\right)= 8\pi
s_{ab}\left(l\right)\,,
\end{equation}
where $l$ is the proper distance
perpendicular to the shell as already seen,
the quantity
$\Omega_{ab}(l)$ is a
geometric quantity, and 
$s_{ab}(l)$ is
the
stress-energy tensor of the double layer that naturally
arises in this junction condition, both quantities having a
distribution character. 
$\Omega_{ab}(l)$
is defined through
\begin{equation}
\label{Omegaabdefinition}
\int_{\mathcal V}
\Omega_{ab}(l) Y^{ab} d{\mathcal V}=-
\int_\Sigma h_{ab}\left[R\right]
n^c\nabla_cY^{ab}d\Sigma \,, 
\end{equation}
with $Y^{ab}$ being an arbitrary
spacetime test tensor function.
Note that the geometric quantity
distribution $\Omega_{ab}(l)$ and, by inheritance, the stress-energy
tensor distribution $s_{ab}(l)$, are defined through the spacetime test
tensorial function $Y^{ab}$ and, although this function is quite
arbitrary, having the sole constraint of being defined in a region of
compact support, the distributional double layer quantity
$\Omega_{ab}(l)$, and so $s_{ab}(l)$, is unique.
The solution  for
$\Omega_{ab}(l)$ of Eq.~\eqref{Omegaabdefinition}
is
\begin{equation}
\label{Omegaabsolution}
\Omega_{ab}(l)= h_{ab}\left[R\right]
n^c\nabla_c\delta\left(l\right) \,, 
\end{equation}
noting that $h_{ab}\left[R\right]$
is a quantity defined on the layer itself, and with the $l$ dependence
of $\Omega_{ab}(l)$ being in the derivative of the delta function
along $l$, i.e., $n^c\nabla_c\delta\left(l\right)$.
To prove that $\Omega_{ab}(l)$ as given in
Eq.~\eqref{Omegaabsolution}
is indeed the sought for solution, we
separate the integral in the spacetime volume ${\mathcal V}$ into an
integral in $\Sigma$ iterated with an integral in $l$ and write
$d{\mathcal V}=d\Sigma dl$.  Then, $\int_{\mathcal V} \Omega_{ab}(l)
Y^{ab} d{\mathcal V}= \int_\Sigma\int_l \Omega_{ab}(l) Y^{ab} dl
d\Sigma $. Assuming $\Omega_{ab}(l)= h_{ab}\left[R\right]
n^c\nabla_c\delta\left(l\right)$ we have
$
\int_\Sigma\int_l \Omega_{ab}(l) Y^{ab} dl d\Sigma=
\int_\Sigma\int_l \left[ h_{ab}\left[R\right]
n^c\nabla_c\delta\left(l\right)\right]
Y^{ab} dl d\Sigma
$. The integral in the right hand side of
the equality can be
performed  by parts,
i.e.,
$
\int_\Sigma\int_l \left[ h_{ab}\left[R\right]
n^c\nabla_c\delta\left(l\right)\right]
Y^{ab} dl d\Sigma
=
\int_\Sigma\int_l
n^c\partial_c\left(
h_{ab}\left[R\right]
\delta\left(l\right)Y^{ab}\right) dl d\Sigma
-
\int_\Sigma\int_l
n^c\nabla_c\left(h_{ab}\left[R\right]
Y^{ab}\right)\delta\left(l\right)
dl d\Sigma$, where in the first part the
covariant derivative is changed into a partial derivative because its
argument is a scalar.  The first part can be found by performing the
integral in $l$ to give the following result $\int_\Sigma\int_l
n^c\partial_c\left(h_{ab}\left[R\right]
\delta\left(l\right)Y^{ab}\right) dl d\Sigma =\int_\Sigma \left[
h_{ab}\left[R\right] \delta\left(l\right)Y^{ab}|_{l_{\rm min}}^{l_{\rm
max}}\right]d\Sigma $, for some $l_{\rm min}$ and $l_{\rm max}$. But
at $l$ different from 0 the delta function is zero and the tensor test
function $Y^{ab}$ has compact support.  So the first part of the
integral vanishes.  The second part of the integral can be performed
in $l$ directly to give
$-
\int_\Sigma\int_l
n^c\nabla_c\left(h_{ab}\left[R\right]Y^{ab}\right)\delta\left(l\right)
dl d\Sigma=-
\int_\Sigma
n^c\nabla_c\left(h_{ab}\left[R\right]Y^{ab}\right)d\Sigma
=-\int_\Sigma
h_{ab}\left[R\right]n^c\nabla_cY^{ab}d\Sigma
$, where in  the last equality
we have passed $h_{ab}\left[R\right]$ to the outside of the
covariant derivative along $l$ since
$h_{ab}\left[R\right]$ is a tensor defined on the layer. 
Thus, indeed, $\Omega_{ab}(l)= h_{ab}\left[R\right]
n^c\nabla_c\delta\left(l\right)$ is the expression for
$\Omega_{ab}(l)$.
Now, substituting $\Omega_{ab}\left(l\right)$
given in Eq.~\eqref{Omegaabsolution}
into 
the original junction condition 
given in Eq.~\eqref{6thgeneralabapp0}
one finds that the double layer stress-energy tensor
distribution $s_{ab}\left(l\right)$
given by the sixth junction condition of the theory 
can also be written in the form
\begin{equation}
\label{sabapp0}
\epsilon\alpha
h_{ab}\left[R\right]n^c\nabla_c\delta\left(l\right)
=4\pi s_{ab}\left(l\right).
\end{equation}
Note that Eq.~\eqref{sabapp0} is Eq.~\eqref{sab} of the main text.

Thin shell with double layers appear in electric dipole
distributions, and the double layer stress-energy tensor distribution
$s_{ab}(l)$ of Eq.~(\ref{sabapp0}) has similarities to those types of
distributions.  The dipole expression of Eq.~\eqref{sabapp0},
which has its origins uniquely from the quadratic term of $f(R)$
theories, here originates
from the term $\alpha R^2$, comes from the coupling
constant $\alpha$ and a nonzero discontinuous
rise of the curvature $R$.  Since in this
gravitational setting, and indeed in gravitation in general, there are
no different types of charges, there is no immediate interpretation, in
physical terms, for the emergence of dipole double layer distributions
in such gravity theories.  In general relativity, thin shell
spacetimes with nonzero $[R]$ exist, but Eq.~(\ref{sab}) does not
manifest itself, since $\alpha = 0$ for general relativity.

\section{Formulas and definitions for the double layer wormhole in
quadratic $R+\alpha R^2$ gravity of Sec.~\ref{sec:worms} with general
exponents $\gamma$ and $\beta$ in the metric functions
of the interior line element}
\label{formulasanddefinitions}

A quite general line element for the interior wormhole region can be
set up by choosing as redshift and shape
functions the following expressions
$\zeta\left(r\right)=\zeta_0\left(\frac{r_0}{r}\right)^\gamma$ and
$b\left(r\right)=r_0\left(\frac{r_0}{r}\right)^\beta$,
respectively,
where $\zeta_0$
is a constant with no units, $r_0$ is the radius of the wormhole
throat, and $\gamma$ and $\beta$ are free exponents.  The line element
for the interior region $\mathcal V^-$ is then
\begin{align}\label{metricconcreteapp}
ds_-^2=-{\rm e}^{\zeta_0\left(\frac{r_0}{r}\right)^\gamma} dt^2+
\frac{dr^2} {1- {\left(\frac{r_0}{r}\right)^{\beta +1}}}
+r^2d\Omega^2\,,\quad\quad\quad r_0\leq r < r_\Sigma\,,
\end{align}
where the $-$ indicates the interior region, and the radial coordinate
$r$ has range $r_0\leq r \leq r_\Sigma$, with $r_\Sigma$ is the radius
of the two separation boundaries, the mouths, one for the upper
domain of the
interior region, the other for the lower
domain of the interior region.  This family
of line elements given by the free exponents $\gamma$ and $\beta$ of
the redshift and shape functions, respectively, have as particular
cases the values $\gamma=1$ and $\beta=1$, which are the values taken
in the main text.  For these values
$\zeta\left(r\right)=\zeta_0\left(\frac{r_0}{r}\right)$,
$b\left(r\right)=r_0\left(\frac{r_0}{r}\right)$, and $ds_-^2 =-{\rm
e}^{\zeta_0\left(\frac{r_0}{r}\right)} dt^2+ \frac{dr^2} {1-
{\left(\frac{r_0}{r}\right)^{2}}} +r^2d\Omega^2$ in the
two domains of the interior region defined
by
$r_0\leq r<r_\Sigma$, which is the line element given in
Eq.~\eqref{metricconcrete}.

The exterior region to be connected to the interior is given by two
Schwarzschild spacetime domains, with line element
\begin{align}\label{schwarzapp}
ds_+^2=-\left(1-\frac{2M}{r}\right)e^{\zeta_e}dt^2+\left(1-
\frac{2M}{r}\right)^{-1}dr^2 + r^2d\Omega^2
\,,\quad\quad
r_\Sigma<
r <\infty\,,
\end{align}
where the $+$ indicates the exterior, $M$ is the spacetime mass, and
$\zeta_e$ is a constant yet to be determined, introduced for later
convenience.  The radial coordinate $r$ has range $r_\Sigma\leq r
<\infty$, where $r_\Sigma$ is the radius of the two separation
boundaries, the mouths, one for the upper domain
of the exterior region, the other for the
lower domain of the exterior
region.  This is the same as the line element of
Eq.~\eqref{schwarz} given now in a general context.

We now have to compute the quantities on which the junction conditions
given in Eqs.~\eqref{junction1}-\eqref{sab} depend, namely,
$h_{00}^-$,
$h_{00}^+$,
$n^{-c}$,
$n^{+c}$,
${{K^-}_{0}}^0$,
${{K^-}_{\theta}}^\theta$,
${K^-}$,
${{K^+}_{0}}^0$,
${{K^+}_{\theta}}^\theta$,
${K^+}$,
$R^{-\Sigma}$,
$R^{+\Sigma}$,
$(\nabla_cR)^{-\Sigma}$, and
$(\nabla_cR)^{+\Sigma}$.
Noting that all
quantities are evaluated at the junction $r_\Sigma$,
and
using the superscripts $-$ and $+$ for the interior and exterior
regions, respectively, 
we have
\begin{equation}
\begin{aligned}
\label{h00}
h_{00}^-=&-e^{\zeta_0\left(\frac{r_0}{r_\Sigma}\right)^\gamma }
\,,
\\
h_{00}^+=&-
\left(1-\frac{2M}{r_\Sigma}\right)e^{\zeta_e}\,,\\
\end{aligned}
\end{equation}
\begin{equation}
\begin{aligned}
\label{normals}
n^{-a}=&(0,\sqrt{1-\left(\frac{r_0}{r_\Sigma}
\right)^{\beta+1}},0,0)\,,
\\
n^{+a}=&(0,\sqrt{1-\frac{2M}{r_\Sigma}},0,0)\,,
\end{aligned}
\end{equation}
\begin{equation}
\begin{aligned}
\label{K00}
{{K^-}_{0}}^0=&-\frac{\gamma\zeta_0}{2r_\Sigma}
\left(\frac{r_0}{r_\Sigma}\right)^\gamma \sqrt{1-
\left(\frac{r_0}{r_\Sigma}\right)^{\beta +1}}\,,
\\
{{K^-}_{\theta}}^\theta=&\frac{1}{r_\Sigma}\sqrt{1-
\left(\frac{r_0}{r_\Sigma}\right)^{\beta+1}}\,,
\\
{{K^-}_{\;\;}}^{\;\;}=&-\frac{1}{2r_\Sigma}\sqrt{1-
\left(\frac{r_0}{r_\Sigma}\right)^{\beta+1}}\left[\gamma
\zeta_0\left(\frac{r_0}{r_\Sigma}\right)^\gamma -4\right],
\\
{{K^+}_{0}}^0=&\frac{M}{r_\Sigma^2}
\sqrt{\frac{r_\Sigma}{r_\Sigma-2M}}\,,
\\
\hskip -2.0cm {{K^+}_{\theta}}^\theta=&\frac{1}{r_\Sigma}
\sqrt{1-\frac{2M}{r_\Sigma}}\,,\;
\\
\hskip -2.0cm {{K^+}_{\;\;}}^{\;\;}=&\frac{2r_\Sigma-3M}{r_\Sigma^2}
\sqrt{\frac{r_\Sigma}{r_\Sigma-2M}}\,,
\end{aligned}
\end{equation}
\begin{equation}
\begin{aligned}
\label{Rscalarcurvature}
R^{-\Sigma}=&\frac{1}{2r_\Sigma^2}\left\{-4\beta
\left(\frac{r_0}{r_\Sigma}\right)^{\beta+1}+
\zeta_0^2\gamma^2\left(\frac{r_0}{r_\Sigma}\right)^{2\gamma}
\left[\left(\frac{r_0}{r_\Sigma}\right)^{\beta+1}-1\right]
\right.
\\
&
\left.
+\zeta_0 \gamma\left(\frac{r_0}{r_\Sigma}\right)^\gamma 
\left[2
\left(1-\gamma\right)+\left(\beta+2\gamma-1\right)\left(
\frac{r_0}{r_\Sigma}\right)^{\beta+1}\right]\right\}\,,
\\
R^{+\Sigma}=& 0\,,
\end{aligned}
\end{equation}
\begin{equation}
\begin{aligned}
\label{derRscalarcurvature}
(\nabla_cR)^{-\Sigma}=&(0,\frac{1}{2r_\Sigma^3}
\left\{4\beta\left(\beta+3\right)\left(
\frac{r_0}{r_\Sigma}\right)^{\beta+1}+\right.
\\
&\left.+\zeta_0\gamma\left(\frac{r_0}{r_\Sigma}\right)^\gamma 
\Bigg[2\left(\gamma^2+\gamma-2\right)-
\left(\gamma+\beta+3\right)\times
\Bigg.\right.
r\\
&
\left.
\times
\left(2\gamma+\beta-1\right)\left(\frac{r_0}{r_\Sigma}
\right)^{\beta+1}\right]+
\zeta_0^2\gamma^2\left(\frac{r_0}{r_\Sigma}
\right)^{2\gamma}\times
\\
&
\left.
\times
\left[2\left(\gamma+1\right)-
\left(2\gamma+\beta+3\right)\left(
\frac{r_0}{r_\Sigma}\right)^{\beta+1}\right]\right\},0,0)\,,
\\
(\nabla_cR)^{+\Sigma}=&(0,0,0,0).
\end{aligned}
\end{equation}
For the case studied in the main text we have to put
$\gamma=1$ and $\beta=1$ in the expressions just given
to find
$h_{00}^-=-e^{\zeta_0\left(\frac{r_0}{r_\Sigma}\right)}$,
$h_{00}^+=-\left(1-\frac{2M}{r_\Sigma}\right)e^{\zeta_e}$,
$n^{-a}=(0,\sqrt{1-\left(\frac{r_0}{r_\Sigma}
\right)^{2}},0,0)$,
$n^{+a}=(0,\sqrt{1-\frac{2M}{r_\Sigma}},0,0)$,
${{K^-}_{0}}^0=-\frac{\zeta_0}{2r_\Sigma}
\left(\frac{r_0}{r_\Sigma}\right) \sqrt{1-
\left(\frac{r_0}{r_\Sigma}\right)^{2}}$,
${{K^-}_{\theta}}^\theta=\frac{1}{r_\Sigma}\sqrt{1-
\left(\frac{r_0}{r_\Sigma}\right)^{2}}$,
${K^-}=-\frac{1}{2r_\Sigma}\sqrt{1-
\left(\frac{r_0}{r_\Sigma}\right)^{2}}\left[
\zeta_0\left(\frac{r_0}{r_\Sigma}\right) -4\right]$,
${{K^+}_{0}}^0=\frac{M}{r_\Sigma^2}
\sqrt{\frac{r_\Sigma}{r_\Sigma-2M}}$,
${{K^+}_{\theta}}^\theta=\frac{1}{r_\Sigma}
\sqrt{1-\frac{2M}{r_\Sigma}}$,
${K^+}=\frac{2r_\Sigma-3M}{r_\Sigma^2}
\sqrt{\frac{r_\Sigma}{r_\Sigma-2M}}$, 
$R^{-\Sigma}=\frac{1}{2r_\Sigma^2}\left\{-4
\left(\frac{r_0}{r_\Sigma}\right)^{2}+
\zeta_0^2\left(\frac{r_0}{r_\Sigma}\right)^{2}
\left[\left(\frac{r_0}{r_\Sigma}\right)^{2}-1\right]
+2\zeta_0 
\left(
\frac{r_0}{r_\Sigma}\right)^{3}\right\}$,
$R^{+\Sigma}= 0$,
$(\nabla_cR)^{-\Sigma}=(0,\frac{1}{2r_\Sigma^3}
\left\{16\left(
\frac{r_0}{r_\Sigma}\right)^{2}-10
\zeta_0\left(\frac{r_0}{r_\Sigma}\right)^3
+
2\zeta_0^2\left(\frac{r_0}{r_\Sigma}
\right)^{2}
\left[2-
3\left(
\frac{r_0}{r_\Sigma}\right)^{2}\right]\right\},0,0)$,
$(\nabla_cR)^{+\Sigma}=(0,0,0,0)$.

In order to shorten the notation, we define
the quantities 
$a(r_\Sigma)$,
$b(r_\Sigma)$,
$c(r_\Sigma)$,
$f(r_\Sigma)$,
$g(r_\Sigma)$,
and $h(r_\Sigma)$.
\noindent
Define  $a(r_\Sigma)$ as $a(r_\Sigma)=\left[K_0^0\right]=
{{K^+}_{0}}^0-
{{K^-}_{0}}^0$, so that
\begin{align}\label{arSigma}
a(r_\Sigma)&=\frac{M}{r_\Sigma^2}
\sqrt{\frac{r_\Sigma}{r_\Sigma-2M}}
+
\frac{\gamma\zeta_0}{2r_\Sigma}
\left(\frac{r_0}{r_\Sigma}\right)^\gamma \sqrt{1-
\left(\frac{r_0}{r_\Sigma}\right)^{\beta+1}}\,.
\end{align}
For the case studied in the main text we  put
$\gamma=1$ and $\beta=1$ in the expressions just given
to find
$a(r_\Sigma)=\frac{M}{r_\Sigma^2}
\sqrt{\frac{r_\Sigma}{r_\Sigma-2M}}
+
\frac{\zeta_0}{2r_\Sigma}
\left(\frac{r_0}{r_\Sigma}\right) \sqrt{1-
\left(\frac{r_0}{r_\Sigma}\right)^{2}}$.
Define  $b(r_\Sigma)$ as $b(r_\Sigma)=K^{0\,\Sigma}_{0}=
\frac12\left(
{{K^+}_{0}}^0+
{{K^-}_{0}}^0\right)$, so that
\begin{align}\label{brSigma}
b(r_\Sigma)=
\frac{M}{2r_\Sigma^2}
\sqrt{\frac{r_\Sigma}{r_\Sigma-2M}}-
\frac{\gamma\zeta_0}{4r_\Sigma}
\left(\frac{r_0}{r_\Sigma}\right)^\gamma \sqrt{1-
\left(\frac{r_0}{r_\Sigma}\right)^{\beta+1}}\,.
\end{align}
For the case studied in the main text we  put
$\gamma=1$ and $\beta=1$ in the expressions just given
to find
$b(r_\Sigma)=
\frac{M}{2r_\Sigma^2}
\sqrt{\frac{r_\Sigma}{r_\Sigma-2M}}-
\frac{\zeta_0}{4r_\Sigma}
\left(\frac{r_0}{r_\Sigma}\right) \sqrt{1-
\left(\frac{r_0}{r_\Sigma}\right)^{2}}$.
\noindent
Define  $c(r_\Sigma)$ as $c(r_\Sigma)=K^{{}\,\Sigma}_{}=
\frac12\left(
{{K^+}_{}}^{}+
{{K^-}_{}}^{}\right)$, so that
\begin{align}\label{crSigma}
c(r_\Sigma)=
\frac12 \left(\frac{2r_\Sigma-3M}{r_\Sigma^2}\right)
\sqrt{\frac{r_\Sigma}{r_\Sigma-2M}}
-\frac12
\left(\frac{1}{2r_\Sigma}\right)\sqrt{1-
\left(\frac{r_0}{r_\Sigma}\right)^{\beta+1}}\left[\gamma
\zeta_0\left(\frac{r_0}{r_\Sigma}\right)^\gamma -4\right]
\,.
\end{align}
For the case studied in the main text we  put
$\gamma=1$ and $\beta=1$ in the expressions just given
to find
$
c(r_\Sigma)=
\frac12 \left(\frac{2r_\Sigma-3M}{r_\Sigma^2}\right)
\sqrt{\frac{r_\Sigma}{r_\Sigma-2M}}
-\frac12
\left(\frac{1}{2r_\Sigma}\right)\sqrt{1-
\left(\frac{r_0}{r_\Sigma}\right)^{2}}\left[
\zeta_0\left(\frac{r_0}{r_\Sigma}\right) -4\right]
$.
\noindent
Define  $f(r_\Sigma)$ as $f(r_\Sigma)=R_\Sigma=
\frac12\left(R^{+\Sigma}+R^{-\Sigma}\right)$, so that
\begin{equation}
\begin{aligned}
\label{frSigma}
f(r_\Sigma)=&
\frac{1}{4r_\Sigma^2}\left\{-4\beta
\left(\frac{r_0}{r_\Sigma}\right)^{\beta+1}+
\zeta_0^2\gamma^2\left(\frac{r_0}{r_\Sigma}\right)^{2\gamma}
\left[\left(\frac{r_0}{r_\Sigma}\right)^{\beta+1}-1\right]+
\right.
\\
&
\left.
\zeta_0 \gamma\left(\frac{r_0}{r_\Sigma}\right)^\gamma
\left[2
\left(1-\gamma\right)+\left(\beta+2\gamma-1\right)\left(
\frac{r_0}{r_\Sigma}\right)^{\beta+1}\right] 
\right\}\,.
\end{aligned}
\end{equation}
For the case studied in the main text we  put
$\gamma=1$ and $\beta=1$ in the expressions just given
to find
$f(r_\Sigma)=\frac{1}{4r_\Sigma^2}\left\{-4
\left(\frac{r_0}{r_\Sigma}\right)^{2}+
\zeta_0^2\left(\frac{r_0}{r_\Sigma}\right)^{2}
\left[\left(\frac{r_0}{r_\Sigma}\right)^{2}-1\right]
+2\zeta_0 
\left(
\frac{r_0}{r_\Sigma}\right)^{3}\right\}$.
\noindent
Define  $g(r_\Sigma)$ as $g(r_\Sigma)=\left[ R\right]=
R^{+\Sigma}-R^{-\Sigma}$, so that
\begin{equation}
\begin{aligned}
\label{grSigma}
g(r_\Sigma)=
&-\frac{1}{2r_\Sigma^2}\left\{-4\beta
\left(\frac{r_0}{r_\Sigma}\right)^{\beta+1}+
\zeta_0^2\gamma^2\left(\frac{r_0}{r_\Sigma}\right)^{2\gamma}
\left[\left(\frac{r_0}{r_\Sigma}\right)^{\beta+1}-1\right]
+
\right.
\\
&
\left.
\zeta_0 \gamma\left(\frac{r_0}{r_\Sigma}\right)^\gamma
\left[2
\left(1-\gamma\right)+\left(\beta+2\gamma-1\right)\left(
\frac{r_0}{r_\Sigma}\right)^{\beta+1}\right]\right\}
\,.
\end{aligned}
\end{equation}
For the case studied in the main text we  put
$\gamma=1$ and $\beta=1$ in the expressions just given
to find
$g(r_\Sigma)=-\frac{1}{2r_\Sigma^2}\left\{-4
\left(\frac{r_0}{r_\Sigma}\right)^{2}+
\zeta_0^2\left(\frac{r_0}{r_\Sigma}\right)^{2}
\left[\left(\frac{r_0}{r_\Sigma}\right)^{2}-1\right]
+2\zeta_0 
\left(
\frac{r_0}{r_\Sigma}\right)^{3}\right\}$.
\noindent
Define  $h(r_\Sigma)$ as $h(r_\Sigma)=n^c
\left[\nabla_c R\right]
=n^{+c}(\nabla_cR)^{+\Sigma}-
n^{-c}(\nabla_cR)^{-\Sigma}$, so that
\begin{equation}
\begin{aligned}
\label{hrSigma}
h(r_\Sigma)=&-
\frac{1}{2r_\Sigma^3
\sqrt{1-\left(\frac{r_0}{r_\Sigma}
\right)^{\beta+1}}
}
\left\{4\beta\left(\beta+3\right)\left(
\frac{r_0}{r_\Sigma}\right)^{\beta+1}+
+\zeta_0 \gamma\left(\frac{r_0}{r_\Sigma}\right)^\gamma
\times\right.
\\
&\times\left.
\Bigg[2\left(\gamma^2+\gamma-2\right)-\left(\gamma+\beta+3\right)
\left(2\gamma+\beta-1\right)\left(\frac{r_0}{r_\Sigma}
\right)^{\beta+1}\right]
\\
&\left.
+\zeta_0^2\gamma^2\left(\frac{r_0}{r_\Sigma}
\right)^{2\gamma}\left[2\left(\gamma+1\right)-
\left(2\gamma+\beta+3\right)\left(
\frac{r_0}{r_\Sigma}\right)^{\beta+1}\right]\right\}
\,.
\end{aligned}
\end{equation}
For the case studied in the main text we  put
$\gamma=1$ and $\beta=1$ in the expressions just given
to find
$h(r_\Sigma)=-
\frac{1}{2r_\Sigma^3
\sqrt{1-\left(\frac{r_0}{r_\Sigma}
\right)^{2}}
}\frac{1}{2r_\Sigma^3}
\left\{16\left(
\frac{r_0}{r_\Sigma}\right)^{2}-10
\zeta_0\left(\frac{r_0}{r_\Sigma}\right)^3
+
2\zeta_0^2\left(\frac{r_0}{r_\Sigma}
\right)^{2}
\left[2-
3\left(
\frac{r_0}{r_\Sigma}\right)^{2}\right]\right\}
$.

Let us now apply the junction conditions provided
in Eqs.~\eqref{junction1}-\eqref{sab}
to match the interior wormhole solution described by the line element
in Eq.~\eqref{metricconcreteapp} with the respective redshift
and shape functions, to the exterior
vacuum solution described by
the line element in Eq.~\eqref{schwarzapp}.
Since the thin shell is timelike we put $\epsilon=1$
in the junction conditions.
The first junction condition, i.e., Eq.~\eqref{junction1},
gives
\begin{equation}\label{setzetaeapp}
e^{\zeta_0\left(\frac{r_0}{r_\Sigma}\right)^\gamma }=
\left(1-\frac{2M}{r_\Sigma}\right)e^{\zeta_e}\,,
\quad\quad\quad\quad r = r_\Sigma
\,.
\end{equation}
For the case studied in the main text we  put
$\gamma=1$ and $\beta=1$ in this expression
to find $e^{\zeta_0\left(\frac{r_0}{r_\Sigma}\right) }=
\left(1-\frac{2M}{r_\Sigma}\right)e^{\zeta_e}$.
Note that $r_\Sigma$, the radius at which the junction between the two
line elements is performed, is set by the second junction
condition. 
The second junction condition, i.e.,
Eq.~\eqref{junction2},  $\left[K\right]=0$,
gives
\begin{equation}\label{setrsigmaapp}
\frac{2r_\Sigma-3M}{r_\Sigma^2\sqrt{1-\frac{2M}{r_\Sigma}}}=
\frac{1}{2r_\Sigma}\sqrt{1-
\left(\frac{r_0}{r_\Sigma}\right)^{\beta +1}}
\left[4-\gamma\zeta_0\left(\frac{r_0}{r_\Sigma}\right)^\gamma \right]
\,,
\quad\quad\quad\quad r = r_\Sigma\,.
\end{equation}
For each particular blend of parameters
$\zeta_0$,
$\gamma$, and $\beta$,
one has to solve
Eq.~\eqref{setrsigmaapp} for $r_\Sigma$ and the matching
between the two metrics must be performed
there, at $r=r_\Sigma$. One must then  
put the value of $r_\Sigma$ into Eq.~\eqref{setzetaeapp} and, for each
blend of parameters, obtain the corresponding value of
$\zeta_e$.
For the case studied in the main text we  put
$\gamma=1$ and $\beta=1$ in this expression
to find
$\frac{2r_\Sigma-3M}{r_\Sigma^2\sqrt{1-\frac{2M}{r_\Sigma}}}=
\frac{1}{2r_\Sigma}\sqrt{1-
\left(\frac{r_0}{r_\Sigma}\right)^{2}}
\left[4-\zeta_0\left(\frac{r_0}{r_\Sigma}\right) \right]$.
The third junction condition, i.e., Eq.~\eqref{stressts}, gives
for a stress-energy tensor
of the form
${S_{\alpha}}^\beta=\text{diag}\left(-\sigma,p,p\right)$
the following expressions
\begin{equation}
\begin{aligned}
\sigma =& \frac{1}{8\pi}
\Bigg[ \;a(r_\Sigma)
\Big(1+2\alpha f(r_\Sigma)\Big)+
\;2\alpha\,\Big( b(r_\Sigma)g(r_\Sigma)-h(r_\Sigma)\Big)
\Bigg]\,,
\quad\quad\quad\quad r = r_\Sigma\,,
\\
p =& \frac{1}{8\pi}
\Bigg[ \frac12 a(r_\Sigma)
\Big(1+2\alpha f(r_\Sigma)\Big)+
\alpha\Big( b(r_\Sigma)g(r_\Sigma)+2h(r_\Sigma)\Big)
\Bigg]\,,
\quad\quad\quad\quad r = r_\Sigma\,,
\label{pressapp}
\end{aligned}
\end{equation}
where 
$a(r_\Sigma)$,
$b(r_\Sigma)$,
$f(r_\Sigma)$,
$g(r_\Sigma)$, and
$h(r_\Sigma)$,
have been previously defined.
For the case studied in the main text we  put
$\gamma=1$ and $\beta=1$ in these quantities
as given above.
The fourth junction condition, i.e.,
Eq.~\eqref{s}, yields
\begin{equation}\label{snewapp}
 P=\frac{\alpha}{4\pi}
 c(r_\Sigma)
g(r_\Sigma)\,,
\quad\quad\quad\quad r = r_\Sigma
\,,
\end{equation}
where $c(r_\Sigma)$ and $g(r_\Sigma)$
have been previously defined, with
$P$ measuring the external normal pressure
supported on $\Sigma$. The term that appears in
the energy-momentum tensor $T_{ab}$
related to the fourth junction condition is $Pn_an_b$,
with the normal
being $n_{-a}=(0,
\left(
1-\left(\frac{r_0}{r_\Sigma}
\right)^{\beta+1}
\right)^{-\frac12}
,0,0)$ and
$n_{+a}=(0,
\left(
1-\frac{2M}{r_\Sigma}
\right)^{-\frac12}
,0,0)$.
The normals  $n_{-a}$ and  $n_{+a}$
are the same at $r_\Sigma$, only
written in different coordinate systems,
indeed $n_{\mp a}
n^{\mp a}=1$ and they are applied at the same point.
For the case studied in the main text we  put
$\gamma=1$ and $\beta=1$ in  $c(r_\Sigma)$ and $g(r_\Sigma)$
as done above.
The fifth junction condition, i.e.,
Eq.~\eqref{sa},
is a condition on the external flux momentum $F_a$,
yielding $F_a=0$, and so its projected part
onto the hypersurface $\Sigma$ satisfies, accordingly,
\begin{equation}\label{sanewapp}
F_\alpha=0\,,
\quad\quad\quad\quad r = r_\Sigma\,.
\end{equation}
The full term that appears in
the energy-momentum tensor $T_{ab}$
related to the fifth junction condition 
is $F_{(a}n_{b)}$ and, since $F_a=0$,
we have consequently that $F_{(a}n_{b)}=0$.
For the case studied in the main text putting
$\gamma=1$ and $\beta=1$ does not change the expression.
The sixth junction condition, i.e., Eq.~\eqref{sab},
gives
for a double layer stress-energy tensor
of the form
${s_{\alpha}}^\beta(l)=\text{diag}\left(\bar
\sigma,\bar p,\bar p\right)n^c\nabla_c\delta\left(l\right)$
the following
\begin{align}\label{6thgenexampleapp}
\bar\sigma=-\frac{\alpha}{4\pi}g(r_\Sigma)
\,,\quad\quad
p&=\;\;\;\frac{\alpha}{4\pi}g(r_\Sigma)\,,
\quad\quad\quad\quad r = r_\Sigma
\,,
\end{align}
where 
$g(r_\Sigma)$
has been previously defined.
For the case studied in the main text we  put
$\gamma=1$ and $\beta=1$ in  $g(r_\Sigma)$ 
as done above.
All the six junction conditions
hold for the two mouths of the wormhole, one
in the upper domain, the other in the lower domain.

Following an identical procedure, a great
number of
different
solutions realizing these specifications
might be composed for
various choices of the parameters.
The formulas above, reduce to the
formulas of the main text for $\beta=1$ and $\gamma=1$.

\section{Violation of the NEC at the wormhole throat in general
relativity as a comparison with the nonviolation of the NEC at the
wormhole throat in the quadratic $R+\alpha R^2$ gravity of
Sec.~\ref{sec:worms}}
\label{necgeneralrelativity}

\begin{figure}[h!]
\includegraphics[scale=0.8]{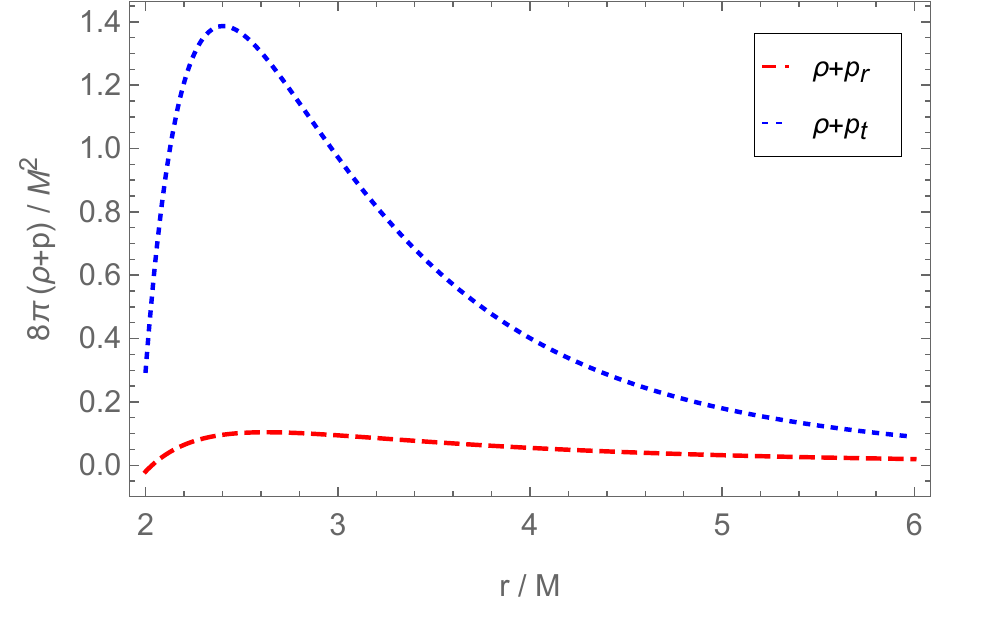}
\caption{Plots of $\rho+p_r$ and $\rho+p_t$ as functions of the radius
$r$ for a wormhole in general relativity, i.e., for $\alpha=0$, and
for $\zeta_0=-60$ and $r_0=2M$, where $M$ is a mass parameter. Since
$\rho+p_r<0$, in general relativity the NEC is violated at the throat,
$r=r_0$.  The quantities plotted are normalized to $M$.
}
\label{fig:necgr}
\end{figure}

In Sec.~\ref{sec:worms}, specifically, in Sec.~\ref{sec:necthroat}, we
have obtained a solution for the interior region of a traversable
wormhole characterized by a distribution that satisfies the matter NEC
at the throat $r_0$.  That the matter NEC at the throat and the
flaring out condition are both simultaneously satisfied is only
possible in the context of modified theories of gravity, which means
$\alpha\neq 0$ in the case presented here.  An instance of it is shown
in Fig.~\ref{fig:nec}.
On the other hand, in general relativity, for
which $\alpha=0$, the NEC is obligatorily violated at the throat.
General relativity is obtained from Eq.~\eqref{field}
by putting $\alpha=0$.  In Fig.~\ref{fig:necgr},
the combinations $\rho+p_r$ and $\rho+p_t$ of the solutions given in
Eq.~\eqref{rhoprpt} with $\alpha=0$ are plotted as functions of $r$
with the other parameters being the same as in Fig.~\ref{fig:nec},
namely, $\zeta_0=-60$, and $r_0=2M$.  From the figure one sees that
$\rho+p_r<0$ and $\rho+p_t>0$, and so in this instance the NEC is
violated at the throat, as it is known to be the case for all wormhole
solutions in general relativity.

\end{document}